\documentclass[mypaper,7pt,twoside]{CoAst}
\usepackage{epsf,graphicx,fancyhdr,url}
\renewcommand{\chaptermark}[1]{\markboth{#1}{}}

\renewcommand{\headrulewidth}{0pt}
\def\setpubdate{will be inserted during the production process}
\newcommand{\kopf}{\small\itshape 
Comm. in Asteroseismology \\ 
Volume No. and date \setpubdate \\
\copyright~Austrian Academy of Sciences}

\def\rfr{\smallskip\par\noindent \hangindent=7truemm \hangafter=1}

\newcommand{\Authors}[1]{\begin{center}\normalsize\bf\sf #1 \end{center}}

\renewcommand{\author}[1]{\begin{center}\normalsize\bf\sf #1 \end{center}}
\newcommand{\Address}[1]{\begin{center}\small\sf #1 \end{center}}

\newcommand{\Objects}[1]{{\vspace{2mm}\small \noindent  \hspace*{0mm}Individual Objects: } \small #1 \normalsize}

\renewenvironment{abstract}{\section*{Abstract}\normalsize\sf}{}
\newcommand{\References}[1]{\begin{flushleft}{\large References\\}\vspace*{2mm}\small #1 \end{flushleft}}

\newcommand{\chapterCoAstfoot}[3]
{\chapter[\sf\normalsize #1\\ 
\footnotesize \hspace*{5mm}by #3 \sf\normalsize][]
{#2\\}%Produces main paper title
\rhead[\fancyplain{}{\sf\footnotesize \center{#1}}]{\fancyplain{}{\sffamily\thepage}}
\lhead[\fancyplain{\kopf}{\sffamily\thepage}]{\fancyplain{\kopf}{\sf\footnotesize \center{#2}}}}

\newcommand{\figureDSSN}[5]{\begin{figure}[#4]
\centering
\includegraphics*[#5]{#1}
\caption{#2}
\label{#3}
\end{figure}}

\renewcommand{\chaptername}{}

\newcommand{\acknowledgments}[1]{\vspace*{5mm}\noindent  \textbf{Acknowledgments.} #1}

\newcounter{subfigure}

\long\def\symbolfootnote[#1]#2{\begingroup%
\def\thefootnote{\fnsymbol{footnote}}\footnote[#1]{#2}\endgroup}

\def\rfr{\smallskip\par\noindent
        \hangindent=7truemm
        \hangafter=1}

\begin{document}
\sf

\chapterCoAstfoot{A spectroscopic study of the hybrid pulsator $\gamma$ Pegasi}{A spectroscopic study of the hybrid pulsator $\gamma$ Pegasi\textsuperscript{\textasteriskcentered}\protect\symbolfootnote[0]{\hspace*{-5mm}\textsuperscript{\textasteriskcentered}
{\bf Based on observations obtained at Vainu Bappu Observatory, Kavalur, India.}}}
{C. P.\,Pandey, T.\,Morel, M.\, Briquet, et al.}
\Authors{C. P.\,Pandey$^{1,2}$, T.\,Morel$^3$, M.\, Briquet$^4$, \\
K.\,Jayakumar$^5$, S.\,Bisht$^1$, and B. B.\,Sanwal$^2$} 
\Address{$^1$ Department of Physics, Kumaun University Nainital - 263002, India.\\
$^2$ Aryabhatta Research Institute of Observational Sciences, Nainital - 263129, India. \\
$^3$ Institut d'Astrophysique et de G\'eophysique, Universit\'e de Li\`ege, All\'ee du 6 Ao\^ut, B\^at. B5c, 4000 Li\`ege, Belgium.\\
$^4$ Katholieke Universiteit Leuven, Departement Natuurkunde en Sterrenkunde, Instituut voor Sterrenkunde, Celestijnenlaan 200D, B-3001 Leuven, Belgium.\\
$^5$ Vainu Bappu Observatory, Indian Institute of Astrophysics, Kavalur - 635701, India.}

\noindent
\begin{abstract}
The recent detection of both pressure and high-order gravity modes in the classical B-type pulsator $\gamma$ Pegasi offers promising prospects for probing its internal structure through seismic studies. To aid further modelling of this star, we present the results of a detailed NLTE abundance analysis based on a large number of time-resolved, high-quality spectra. A chemical composition typical of nearby B-type stars is found. The hybrid nature of this star is consistent with its location in the overlapping region of the instability strips for $\beta$ Cephei and slowly pulsating B stars computed using OP opacity tables, although OPAL calculations may also be compatible with the observations once the uncertainties in the stellar parameters and the current limitations of the stability calculations are taken into account. The two known frequencies $f_1 = 6.58974$ and $f_2 = 0.68241$~d$^{-1}$ are detected in the spectroscopic time series. A mode identification is attempted for the low-frequency signal, which can be associated to a high-order $g$-mode. Finally, we re-assess the binary status of $\gamma$ Peg and find no evidence for variations that can be ascribed to orbital motion, contrary to previous claims in the literature. 
\end{abstract}

\Objects{$\gamma$\, Pegasi.} 

\section*{Introduction}
Of particular relevance for our understanding of the fundamental properties of stars on the upper main sequence are the so-called hybrid B pulsators, which simultaneously exhibit low-order pressure and high-order gravity modes characteristic of $\beta$ Cephei and slowly pulsating B stars (hereafter SPBs), respectively (in a similar vein, some A- and F-type main sequence stars also present ${\gamma}$ Doradus- and ${\delta}$ Scuti-like pulsations; e.g., Grigahc\`ene et al. 2010). A number of hybrid $\beta$ Cephei/SPB pulsating candidates have been identified to date (e.g., De Cat et al. 2007, Pigulski \& Pojma\'nski 2008, Degroote et al. 2009), but their number is expected to grow dramatically in the future as intensive space observations with unprecedented duty cycle and photometric precision are being undertaken (e.g., Balona  et al. 2011). The self-driven excitation of both pressure and gravity modes holds great asteroseismic potential because it offers an opportunity to probe both the stellar envelope and the deep internal layers. Such stars are hence prime targets for in-depth seismic modelling and, although challenging, their study promises to lead to significant advances in our understanding of the internal structure of main-sequence B stars (e.g., internal rotation profile, extent of the convective core; Thoul 2009). 

Of particular interest in this context is the bright B2 IV star $\gamma$ Peg (HR 39, HD 886). Although for long considered as a classical $\beta$ Cephei star, it has recently also been shown to exhibit gravity mode pulsations typical of SPBs, as first shown by Chapellier et al. (2006) using ground-based spectroscopic observations. This result has recently been confirmed and extended by Handler et al. (2009) from high-precision photometry with the {\it MOST} satellite and a coordinated radial-velocity monitoring from the ground, with the detection of eight $\beta$ Cep-like and six SPB-like oscillation modes (see also Handler 2009). With this peculiarity, ${\gamma}$ Peg enters the (so far) restricted club of hybrid pulsators and is expected to play in the future a pivotal role towards a better understanding of the physics of B stars. Modelling this star also has the potential to address even more fundamental issues, such as the reliability of opacity calculations (see the attempts in this direction by Walczak \& Daszy\'nska-Daszkiewicz 2010 or Zdravkov \& Pamyatnykh 2009).

However, the reliability of the results provided by such seismic studies strongly hinges upon an accurate knowledge of both the fundamental parameters of the star under study (e.g., Teff) and its metal mixture. For this reason, an accurate determination of these quantities is of vital importance. In virtue of its brightness and initial status as a prototypical $\beta$ Cephei star, many abundance works have been devoted to $\gamma$ Peg (e.g., the pioneering work of Aller 1949), but the analyses very often relied on non fully line-blanketed model atmospheres or fundamental parameters inferred from photometric calibrations and/or LTE calculations (e.g., Gies \& Lambert 1992, Pintado \& Adelman 1993, Ryans et al. 1996). These limiting assumptions cast some doubts on the reliability of these results. For instance, fitting of the wings of the Balmer lines using LTE synthetic spectra have been shown to systematically lead to an overestimation of the surface gravity (Nieva \& Przybilla 2007). Only a few studies have been conducted from an NLTE perspective (Andrievsky et al. 1999, Korotin et al. 1999a,b, Morel et al. 2006), all but one of them only deriving the CNO abundances.\footnote{Note that the NLTE abundances of Gies \& Lambert (1992) are not based on full NLTE line-formation calculations, but are derived instead from the LTE values assuming theoretical NLTE corrections appropriate for a star with parameters typical of $\gamma$ Peg.} Here we present a detailed NLTE abundance analysis of this star in an effort to re-examine its fundamental properties and its position relative to the theoretical instability domains for SPBs and $\beta$ Cephei stars. We also take advantage of the large number of time-resolved, high-quality spectra collected to re-assess its binarity and to attempt to identify the most prominent modes visible in spectroscopy.

\section*{Observations and data reduction} 
The spectroscopic observations were obtained at Vainu Bappu Observatory (VBO) located in Kavalur (India) using the fiber-fed \'echelle spectrograph attached to the prime focus of the 2.3-m Vainu Bappu telescope (VBT). The wavelength range was 4000-8000 \AA \ for the 2 K $\times$ 4 K CCD detector (spread over 45 orders) and 4200-7000 \AA \ for the 1 K $\times$ 1 K CCD detector (spread over nearly 25 orders with the \'echelle gaps). The resolving power estimated from the arc spectra is $R = \lambda/\Delta \lambda$ $\approx$ 60\,000. Further details regarding the instrumental set up are given in Rao et al. (2005). 

A total of 163 spectra were collected from September 2007 to January 2009 during various observing runs. The journal of observations is presented in Table \ref{tab:observations}. The exposure time ranged from 10 to 15 minutes depending on the sky conditions (i.e., clear sky or sky with thin clouds) and the signal-to-noise ratio was generally above 200. The data were reduced and analysed using standard IRAF (Image Reduction and Analysis Facility)\footnote{{\tt IRAF} is distributed by the National Optical Astronomy Observatories, operated by the Association of Universities for Research in Astronomy, Inc., under cooperative agreement with the National Science Foundation.} tasks. The basic steps of the data reduction included trimming, bias frame subtraction, scattered light removal, flat fielding, order extraction, wavelength calibration (using ThAr lamps) and finally continuum rectification.

\begin{table*}[h]
\caption{Journal of observations.}
\begin{center}
\begin{tabular}{c c|c c} \hline\hline
{Civil Date}&{Number of spectra} & {Civil Date} & {Number of spectra} \\\hline
20 Sep. 07       & 10 & 02 Feb. 08       &  4   \\
28 Sep. 07       & 19 & 03 Jun. 08       &  4   \\
02 Oct. 07       & 24 & 04 Jun. 08       &  4   \\
04 Oct. 07       & 11 & 05 Jun. 08       &  2   \\
05 Oct. 07       & 28 & 24 Aug. 08       &  5   \\
01 Jan. 08       &  3 & 31 Aug. 08       &  3   \\
02 Jan. 08       & 11 & 04 Oct. 08       &  7   \\
03 Jan. 08       &  5 & 05 Oct. 08       & 11   \\
16 Jan. 08       &  3 & 13 Jan. 09       &  3   \\
17 Jan. 08       &  3 & 14 Jan. 09       &  3   \\ \hline
                 &    & Total            & 163  \\ \hline
\end{tabular}
\label{tab:observations}
\end{center}
\end{table*}

\section*{Line-profile variations}
It should be noted that because of different instrumental settings, not all spectral lines were systematically covered during the observations. We therefore focus in the following on the strong spectral lines that were the most extensively observed, namely the Si III triplet between 4552 and 4575 \AA \ and the C II $\lambda\lambda$5143, 5145 doublet, with 80 and 131 exposures, respectively.

\subsection*{Binarity}
Until recently, $\gamma$~Peg was believed to be a spectroscopic binary whose orbital period was, however, disputed ($P_{\rm orb}=$ 370.5 days, Chapellier et al. 2006; $P_{\rm orb} =$ 6.816 days, Harmanec et al. 1979, Butkovskaya \& Plachinda 2007). Contrary to previous authors, Handler et al. (2009) suggested $\gamma$~Peg to be a single star and explained the claimed orbital variations as due to $g$-mode pulsation. 

To re-assess the possible binarity of this object, we compared our radial velocities (RVs) of the C II $\lambda$5145 line with the orbital solution proposed by Chapellier et al. (2006). Our heliocentrically-corrected RVs vary between $-1.5$ and 7 km\,s$^{-1}$ (with a typical accuracy on the individual measurement of $\sim$0.3 km\,s$^{-1}$) mainly as a result of the well-known, dominant radial mode with a period of about 0.15 day (see below). There is no evidence for any long-term trends, despite the fact that our observations cover some phase intervals (most notably around $\phi$$\sim$0.93) where rapid variations amounting to up to $\sim$12 km\,s$^{-1}$ are expected according to the ephemeris of Chapellier et al. (2006). This rules out the possibility that $\gamma$~Peg is an eccentric binary with an orbital period of about one year, confirming the conclusion of Handler et al. (2009). On the other hand, these authors favoured the pulsation interpretation for the $\sim$6.8 days period since they noticed that the frequency 1/6.816 d$^{-1}$ with an amplitude of $\sim$0.8 km\,s$^{-1}$ (Butkovskaya \& Plachinda 2007), is a one-day alias of the frequency 0.8533 d$^{-1}$, which lies well within the domain of $g$-mode frequencies. Finally, our data also do not support the large, abrupt RV changes reported by Butkovskaya \& Plachinda (2007). 

\subsection*{Frequency analysis and mode identification}
To further investigate the line-profile variability in $\gamma$~Peg, we used the software package FAMIAS\footnote{FAMIAS has been developed in the framework of the FP6 European Coordination Action HELAS -- {\tt http://www.helas-eu.org/}} (Zima 2008). We computed the first three velocity moments \mbox{$<v^1>$}, \mbox{$<v^2>$} and \mbox{$<v^3>$} (see Aerts et al. 1992 for a definition of the moments of a line profile) of the Si III triplet and the C II doublet with the aim of performing a frequency analysis. The integration limits for computing the moments were dynamically chosen by sigma clipping to avoid the noisy continuum.  

 In the first moment \mbox{$<v^1>$}, which is the RV placed at average zero, we detected the well-known dominant radial mode, which can readily be seen in the line-profile variations affecting the Si III lines (Fig.\,\ref{fig:lpv}), followed by a one-day alias of the known highest amplitude low frequency mode. Both pulsation modes were first discovered by Chapellier et al. (2006) and afterwards confirmed by Handler et al. (2009), who definitely proved the hybrid $\beta$~Cep/SPB nature of $\gamma$~Peg. Since the frequency values of Handler et al. (2009) are more accurate than ours, we adopted them in our study, i.e., $f_1 = 6.58974$ and $f_2 = 0.68241$~d$^{-1}$. The corresponding amplitudes in our dataset are 3.39 and 0.69 km\,s$^{-1}$ with an error of 0.07 km\,s$^{-1}$, for $f_1$ and $f_2$, respectively.  No additional frequencies could be detected according to the 4 signal-to-noise criterion of Breger et al. (1993). The latter was tested by computing the noise level for different box intervals (between 1 and 10\,d$^{-1}$) centred on the considered frequency.

\figureDSSN{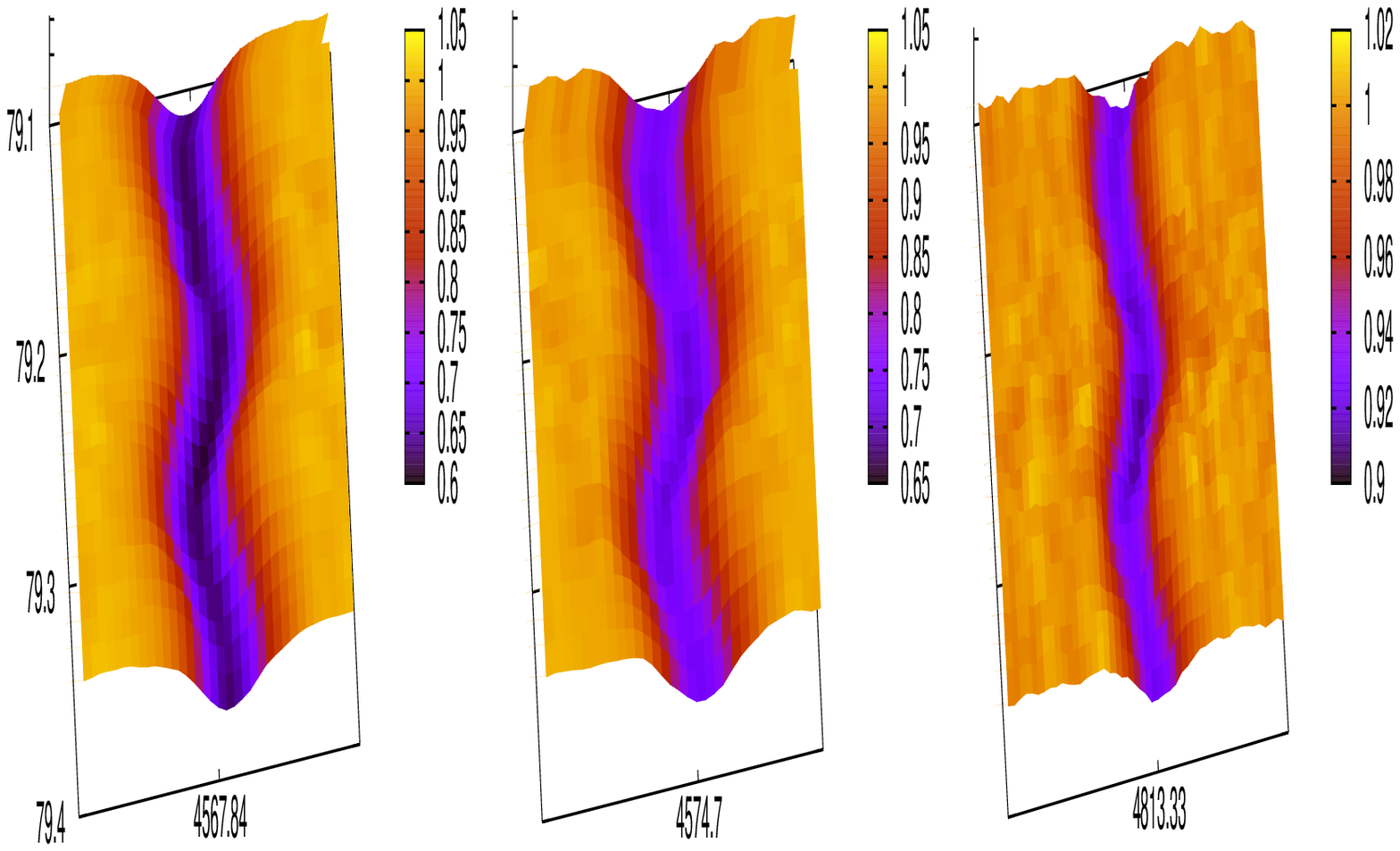}{Examples of line-profile variations affecting Si III $\lambda$4567.8, 4574.7 and 4813.3 on 5 October 2007.}{fig:lpv}{!ht}{clip=0,trim=2cm  15cm 1cm 2cm, angle=0,width=1\textwidth}

\begin{figure*}
\centering
\resizebox{0.88\linewidth}{!}{\rotatebox{-90}{\includegraphics{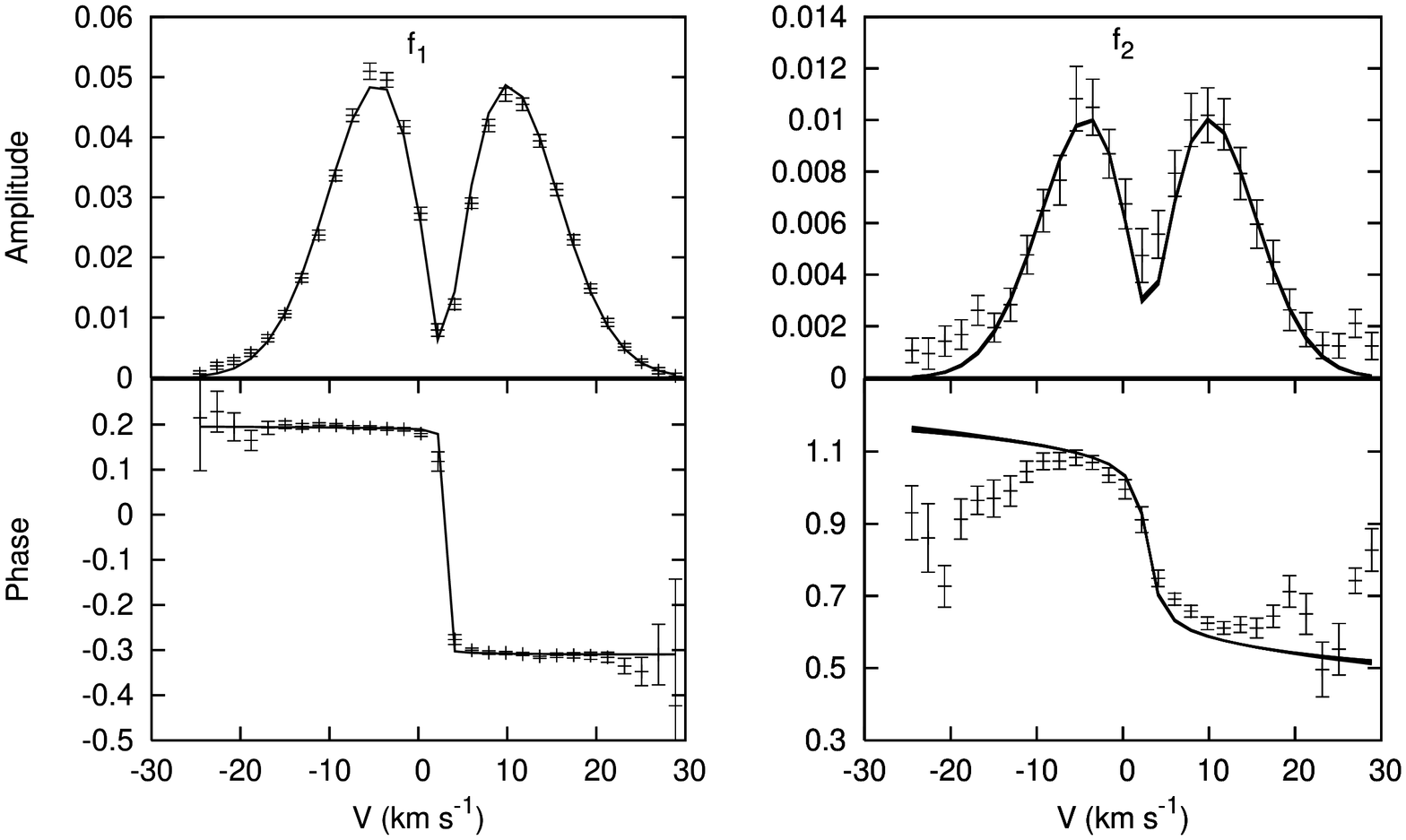}}}
\caption{Amplitude and phase distributions (points with error bars) across the C II $\lambda$5145 \AA \ line, for the frequencies $f_1 = 6.58974$ and $f_2 = 0.68241$~d$^{-1}$. The bestfit model is represented by full lines. For $f_1$, the mode is radial. For $f_2$, the differences between the best solutions listed in Table\,\ref{mi} for ($\ell_2,m_2$)=(1,1), ($2,-1$) and ($2,-2$) are smaller than the thickness of the lines. The amplitudes are expressed in units of continuum and the phases in $\pi$ radians.}
\label{amp_phase}
\end{figure*}

In an attempt to detect other periodicities, we performed a frequency search on the spectra by means of a two-dimensional Fourier analysis. In this way, we again detected $f_1$ and $f_2$ but no additional frequencies. A non-linear multi-periodic leastsquares fit of a sum of sinusoidals is then computed with the Levenberg-Marquardt algorithm   (adapted from Press et al. 2007). This fitting is applied for every bin of the spectrum separately according to the formula $Z+\sum_i A_i \sin\bigl[2\pi (f_i t+\phi_i)\bigr],$ where $ Z$ is the zeropoint, and $ A_i$, $ f_i$, and $ \phi_i$ are the amplitude, frequency, and phase of the $i$-th frequency, respectively. The amplitude and phase distributions across the C II $\lambda$5145 \AA \ line are shown in Fig.\,\ref{amp_phase}. 

\begin{figure} 
\centering
\resizebox{0.85\linewidth}{!}{\rotatebox{-90}{\includegraphics{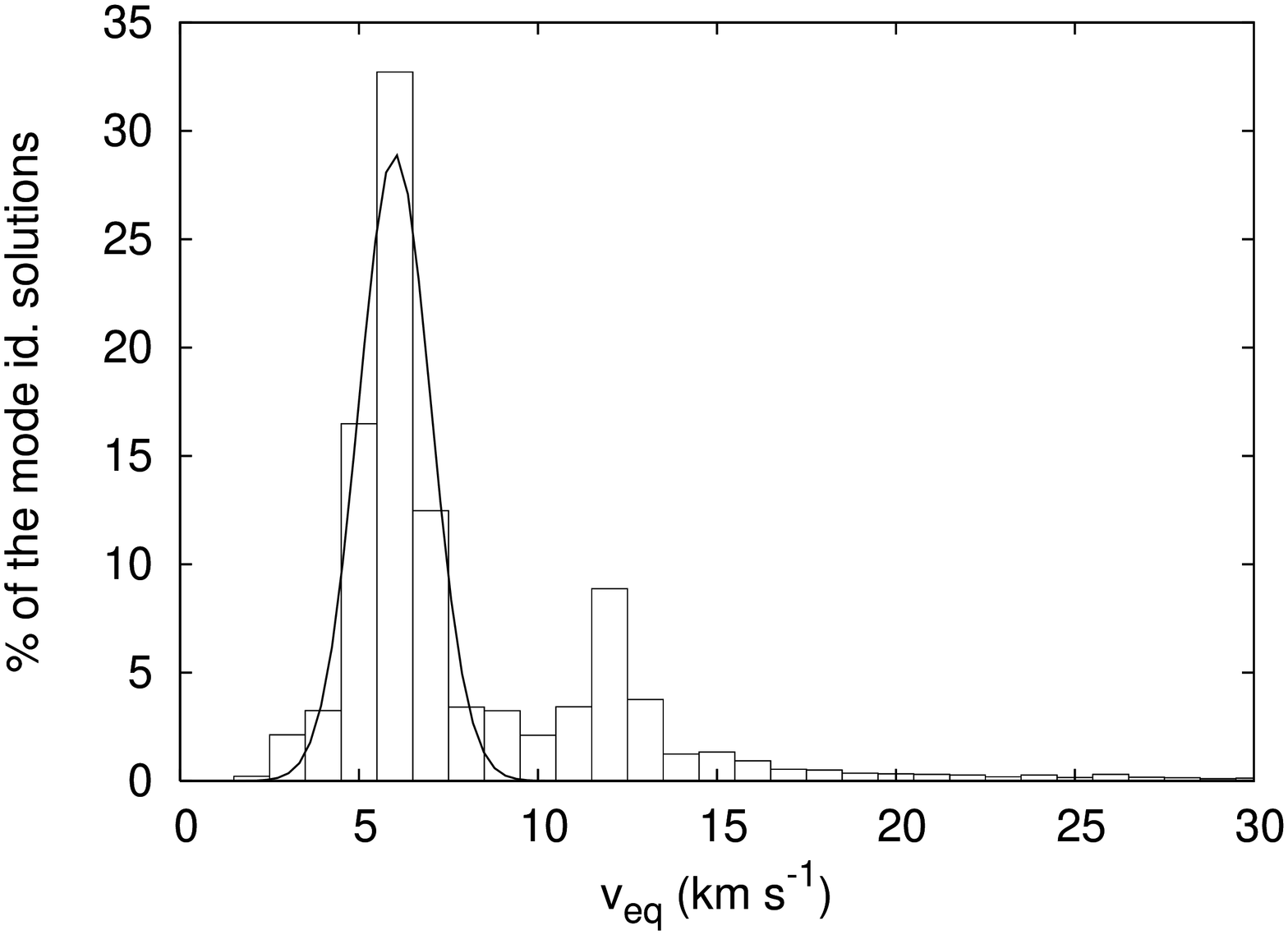}}}
\caption{Histogram for the equatorial rotational velocity derived from the FPF method.}  
\label{histo}  
\end{figure}

To identify the modes associated to $f_1$ and $f_2$, we used the Fourier parameter fit method (FPF method; Zima 2006). The wavenumbers $(\ell,m)$ and other continuous parameters are determined in such a way that the theoretical zeropoint, amplitude and phase values across the profile best fit the observed values. The fitting is carried out by applying genetic optimization routines in a large parameter space. 
From multicolour photometric time series, Handler (2009) unambiguously identified $f_1$ as a radial mode while two possibilities remain for $f_2$, which is a $\ell_2 = 1$ or a $\ell_2 = 2$ mode. Our spectroscopy corroborates that $(\ell_1,m_1) = (0,0)$, as illustrated in Fig.\,\ref{amp_phase}. To identify the values of $(\ell_2,m_2)$, we only allowed $\ell_2$ to be 1 or 2. Such an approach, which consists in adopting the $\ell$-values obtained from photometry, already proved to be successful for spectroscopic mode identification of other $\beta$~Cephei stars (Briquet et al. 2005; Mazumdar et al. 2006; Desmet et al. 2009).
In Table\,\ref{mi} the best parameter combinations are listed for each couple $(\ell_2,m_2)$. The smaller the $\chi^2$ value, the better the solution. The most probable and equally good solutions are ($\ell_2,m_2$) = (1,1), ($2,-2$) or ($2,-1$) (where a positive $m$-value denotes a prograde mode). The best models for the amplitude and phase across the line profile are illustrated in Fig.\,\ref{amp_phase}. A by-product of the mode identification is the derivation of the equatorial  rotational velocity $v_{\rm eq}$. By using the $\chi^2$-values as weights, we constructed a histogram for $v_{\rm eq}$ (see Fig.\,\ref{histo}), as in Desmet et al. (2009). We computed it by considering the solutions with ($\ell_2,m_2$) = (1,1), ($2,-2$) and ($2,-1$). By calculating a weighted mean and standard deviation, we obtained $v_{\rm eq}$ = 6$\pm$1 km\,s$^{-1}$ (a much less probable value is $\sim$12 km\,s$^{-1}$).

\begin{table}
\caption{Mode parameters derived from the FPF mono-mode method for $f_2$. $a_s$ is the surface velocity amplitude (km~s$^{-1}$); $i$ is the stellar inclination angle in degrees; $v\sin i$ is the projected rotational velocity, $\sigma$ is the width of the intrinsic profile, both expressed in km~s$^{-1}$.}
\label{mi}
\begin{center}
\begin{tabular}{cccccc} \hline\hline
$\chi^2$& ($\ell_2,m_2$) & $a_s$ & $i$ & $v\sin i$ & $\sigma$ \\
\hline
3.59 & (1,1)    & 3.9 & 14  & 1.3 & 6.9 \\
3.60 & ($2,-2$) & 2.5 & 25  & 2.6 & 6.5 \\
3.62 & ($2,-1$) & 2.5 & 84  & 6.2 & 6.1 \\
4.77 & (1,0)    & 6.2 & 86  & 5.8 & 5.3 \\
4.89 & (0,0)    & 2.8 & --- & 1.0 & 6.9 \\
4.91 & (2,0)    & 0.4 & 26  & 4.7 & 6.6 \\
5.29 & (2,1)    & 5.1 & 87  & 1.0 & 6.4 \\
5.30 & ($1,-1$) & 0.9 & 63  & 1.0 & 7.1 \\
5.84 & (2,2)    & 7.4 & 15  & 1.0 & 6.1 \\
\hline
\end{tabular}
\end{center}
\end{table}

\section*{Abundance analysis}
\subsection*{Analysis tools}
The atmospheric model calculations were performed under the assumption of LTE, whereas a full NLTE treatment was adopted for the line formation. Such a hybrid approach has been shown to be adequate for early B-type stars on the main sequence (Nieva \& Przybilla 2007). First, the ATLAS9 code\footnote{http://kurucz.harvard.edu/} (Kurucz 1993)  is used to compute the hydrostatic, plane-parallel and fully line-blanketed LTE atmospheric models. Grids with the new opacity distribution functions (ODFs)  assuming the solar abundances of Grevesse \& Sauval (1998) have been used (Castelli \& Kurucz 2004).

Further, in order to obtain the NLTE abundances we made use of the latest version of the NLTE line-formation codes DETAIL/SURFACE (Butler \& Giddings 1985, Giddings 1981). DETAIL provides the solution of the radiative transfer and statistical equilibrium equations, while the emergent spectrum is calculated by SURFACE. The line atomic data are taken from the NIST and VALD databases. Care has been taken to only retain features that are unblended in the relevant temperature range. 

\subsection*{Determination of the atmospheric parameters}
Our abundance analysis is based on the average of a large number of time-resolved spectra (selected to have the highest S/N) to ensure that the parameters and abundances we derive are representative of the values averaged over the whole pulsation cycle.

The effective temperature, $T_{\rm eff}$, was estimated from the Si II/III ionisation balance. We cannot rely upon Si III/IV ionisation balance, as no Si IV lines are visible. On the other hand, in this $T_{\rm eff}$ range no other chemical elements have lines of two adjacent ionisation stages that can be measured. Numerous Fe III lines and two very weak Fe II lines (Fe II $\lambda$5018 and Fe II $\lambda$5169) with an EW in the range 10-15 m\AA \ are present in our spectra. However, constraining $T_{\rm eff}$ from ionisation balance of iron was not possible owing to the rudimentary nature of our Fe II model atom, which only includes 8 levels. In contrast, the detailed treatment of the Fe III ion (264 levels) allows us to reliably estimate the iron abundance from the analysis of the Fe III features. Figure \ref{fig:calibrations} shows examples of calibrations between the Si line ratios and $T_{\rm eff}$. Ten line ratios have been used for the temperature estimation and the dispersion between the various $T_{\rm eff}$ values obtained was considered as representative of the uncertainty in this parameter.

\figureDSSN{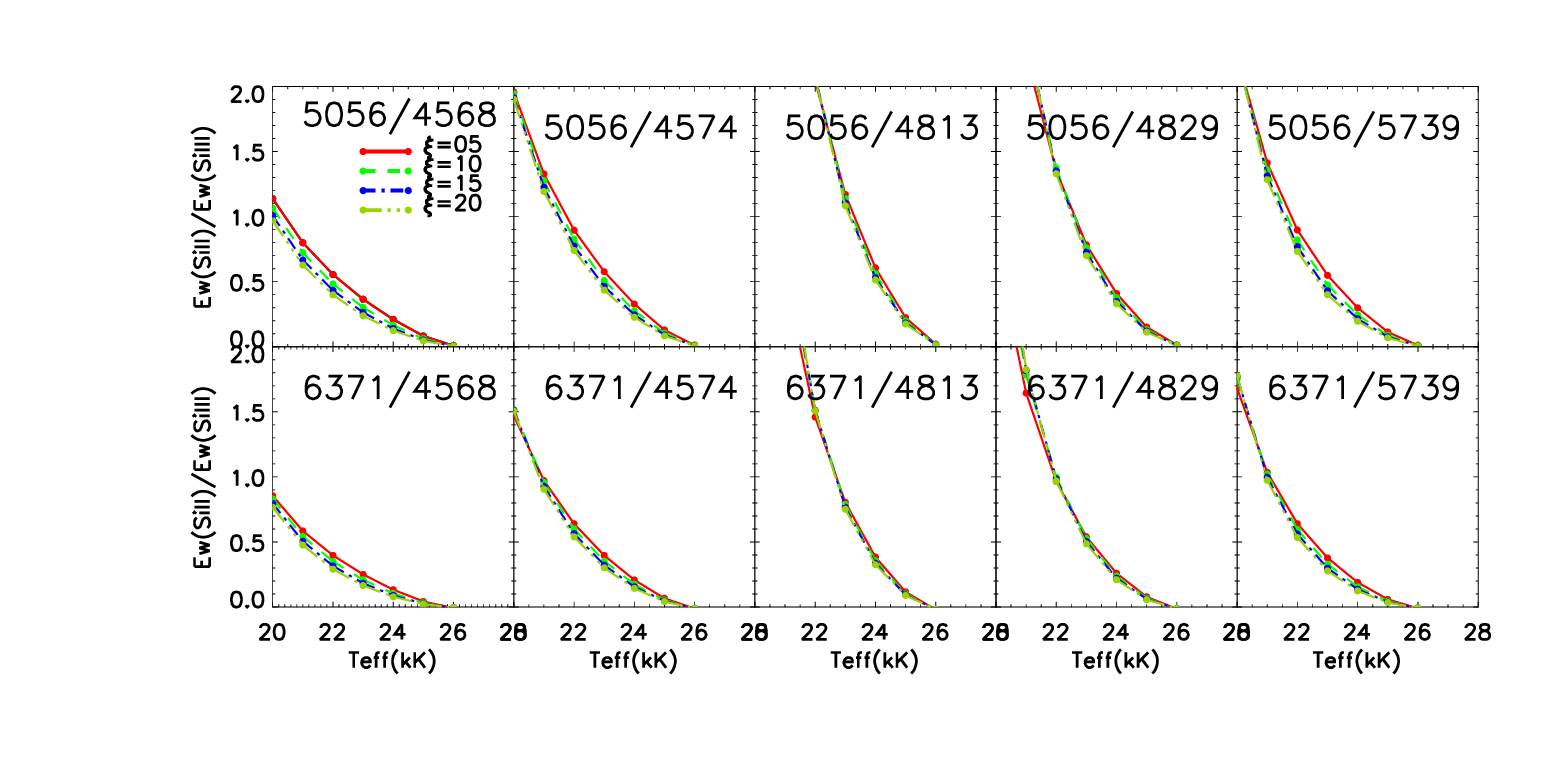}{Examples of calibrations between various Si line ratios and the effective temperature, as a function of the microturbulence {$\xi$}  (red lines: {$\xi$ = 5 km s$^{-1}$}; light green lines: {$\xi$ = 10 km s$^{-1}$}; blue lines: $\xi$ = 15 km s$^{-1}$; dark green lines: {$\xi$ = 20 km s$^{-1}$}). We have adopted $\log g$ = 3.7  and $\xi$ = 5 km s$^{-1}$, as appropriate for this star (see text).}{fig:calibrations}{!ht}{clip,trim= 3cm 1cm 1cm 1cm, angle=0,width=1\textwidth}

Unfortunately, we were unable to estimate the surface gravity, $\log g$, from the fitting of the collisionally-broadened wings of the Balmer lines, as no profiles were completely covered because of \'echelle gaps. Therefore, this parameter was estimated from Str\"omgren {\it uvby$\beta$} photometry based on the calibrations of Castelli (1991). The $c_{1}$ and $\beta$ colour indices were taken from the Catalogue of $uvby\beta$ data of Hauck \& Mermilliod (1998). The observed $c_{1}$ colour index was dereddened to obtain $c_{0}$ by means of empirical calibrations (details can be found in Castelli 1991 and references therein). Slightly different sets of colour indices ($c_{1}$, $\beta$) are available in the literature (Crawford et al. 1971, Sterken \& Jerzykiewicz 1993). The surface gravity was computed using all the possible combinations and the scatter was taken as the typical error bar. The simultaneous estimation of the effective temperature and surface gravity, along with the error limits, is illustrated in Fig.\,\ref{fig:Teff_logg}. 

\figureDSSN{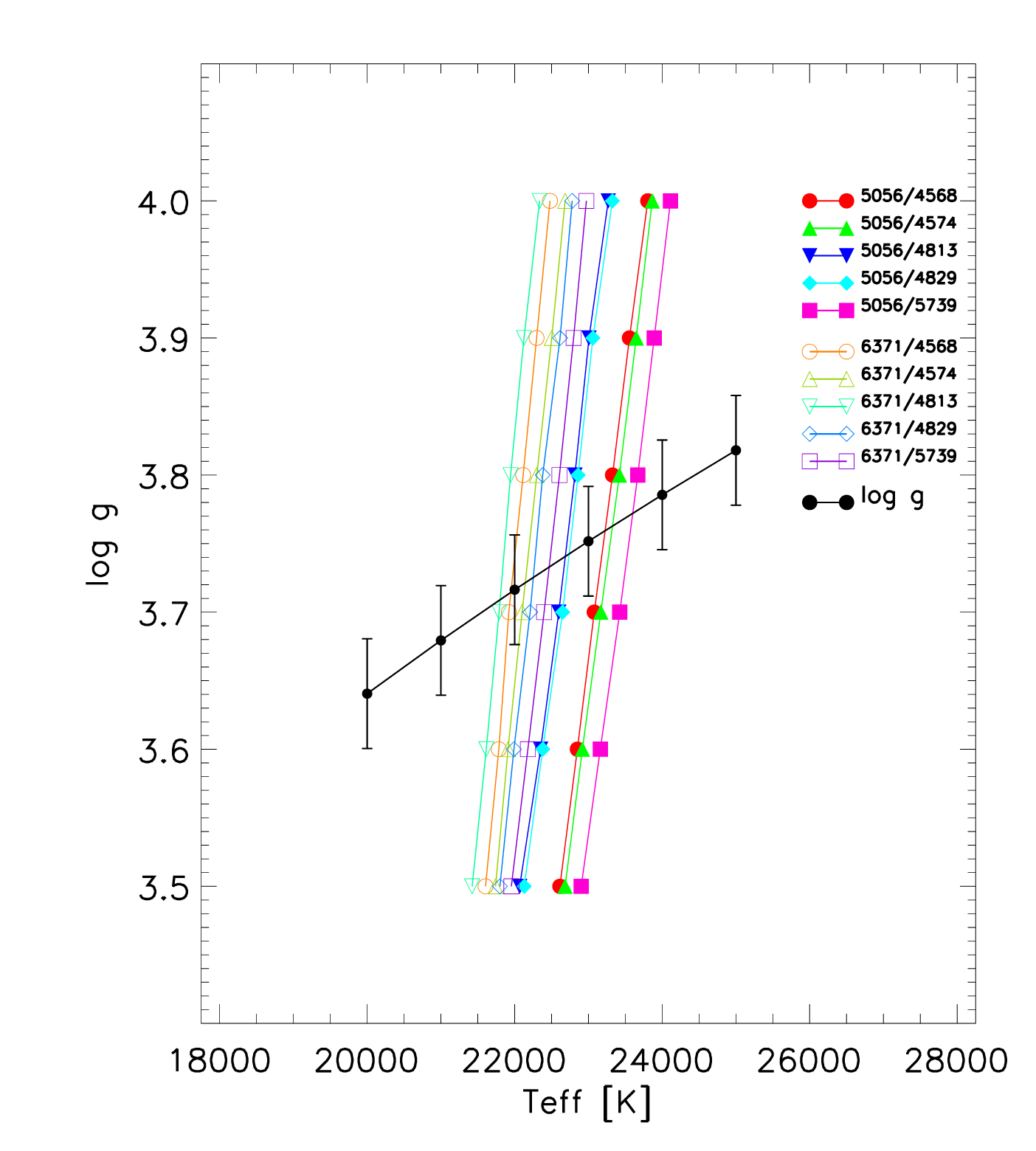} {$T_{\rm eff}$-$\log g$ plane for the simultaneous determination of the effective temperature and surface gravity. The group of nearly vertical lines represents the loci satisfying the Si II/Si III  ionisation equilibria, while the thick solid line represents the $\log g$ determined from Str\"omgren photometry. The middle point of the intersection of the two sets of lines simultaneously provides the value of $T_{\rm eff}$ and $\log g$.} {fig:Teff_logg}{!ht}{clip,angle=0,width=.9\textwidth}

Another important parameter which is required for the abundance analysis is the microturbulent velocity, $\xi$, which was determined from the O II features requiring that the abundances are independent of the line strength (Fig.\,\ref{fig:xi}). The uncertainty in the microturbulence was estimated by varying this parameter until the slope of the $\log \epsilon$(O)-$\log \epsilon$(EW/$\lambda$) differs from zero at the 3$\sigma$ level. 

\figureDSSN{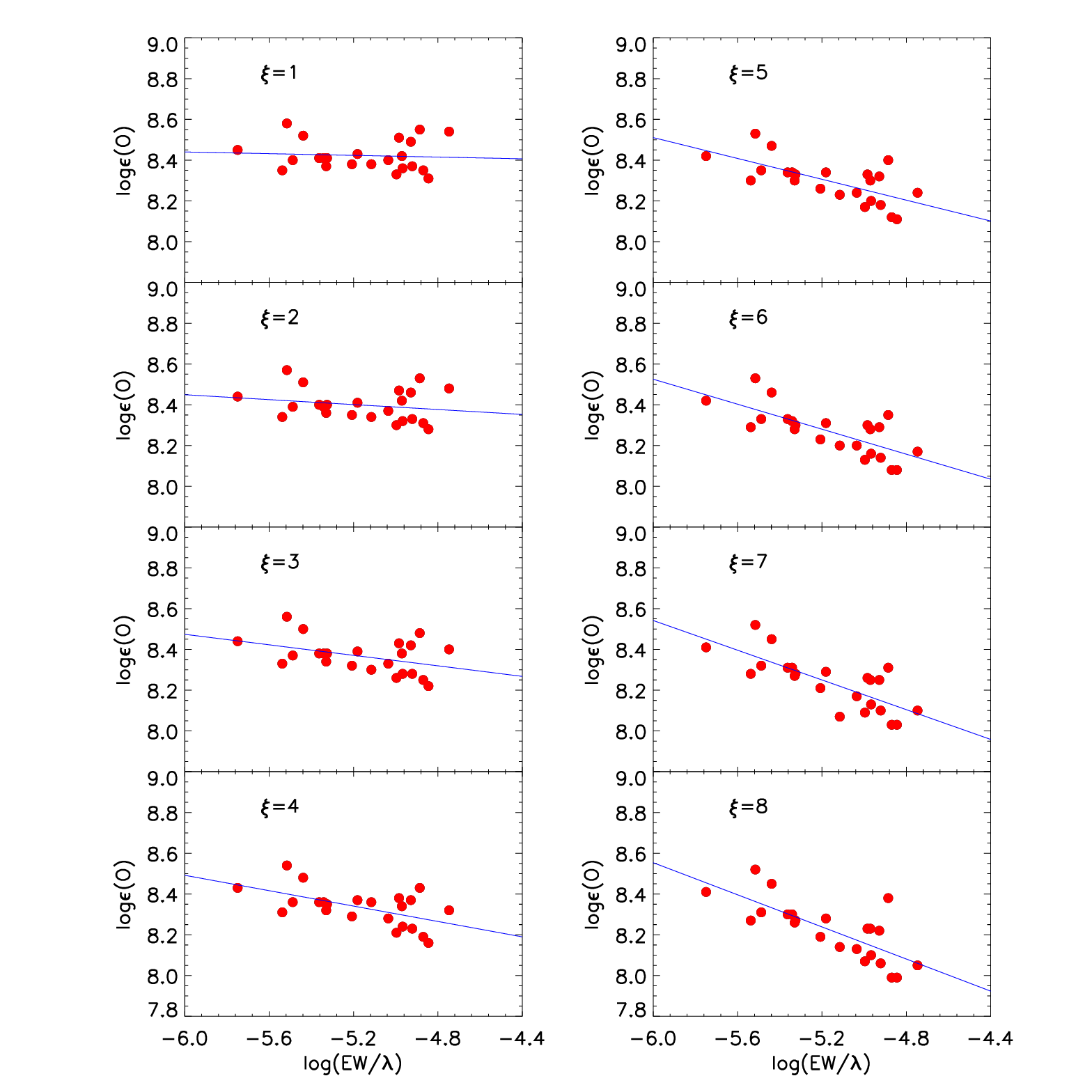}{Estimation of the microturbulent velocity from the O II lines.}{fig:xi}{!ht}{clip,angle=0,width=.9\textwidth, height=.85\textwidth}

Finally, the total amount of line broadening, $v_T$, is determined from line-profile fitting of a number of isolated metallic lines (instrumental broadening was estimated from the calibration lamps). Although rotational broadening clearly dominates over pulsational broadening in this star whose dominant pulsation mode is radial, the fact that this estimate ($v_T$ = 8$\pm$1 km s$^{-1}$) is marginally higher than the equatorial rotational velocity inferred from the moment analysis ($v_{\rm eq}$ = 6$\pm$1 km\,s$^{-1}$) suggests that broadening arising from non-radial pulsations may not be completely negligible. A comparison between the observed and fitted profiles is presented in the Appendix. 

The atmospheric parameters, along with their uncertainties, are presented in Table \ref{tab:parameters}. A detailed comparison with previous results in the literature is deferred to the last section. 

\begin{table*}[h]
\caption{Derived atmospheric parameters for ${\gamma}$ Peg.}
\begin{center}
\begin{tabular}{lc} \hline\hline
$T_{\rm eff}$ (K)        & 22650$\pm$650\\
$\log T_{\rm eff}$       & 4.355$\pm$0.013\\
$\log g$                 & 3.73$\pm$0.08\\
$\xi$      (km s$^{-1}$) & 1${+2 \atop -1}$ \\   
$v_T$ (km s$^{-1}$)      & 8$\pm$1\\
\hline
\end{tabular}
\label{tab:parameters}
\end{center}
\end{table*}

\subsection*{Determination of the elemental abundances}
The NLTE abundances of He, C, N, O, Ne, Mg, Al, Si and Fe have been determined from a classical curve-of-growth analysis using the EWs measured through direct integration. The mean abundances are presented in Table \ref{tab:abundances}, while the results for each transition considered are given in the Appendix (along with the EW measurements). To estimate the abundance uncertainties, we have first calculated the errors arising from the uncertainties in the atmospheric parameters (i.e., $\sigma_{\rm Teff}$, $\sigma_{\rm logg}$ and $\sigma_{\rm \xi}$). They were derived by computing the abundances using models with atmospheric parameters deviating from the nominal values by the uncertainties tabulated in Table \ref{tab:parameters}. We also considered the fact that the determinations of $T_{\rm eff}$ and $\log g$ are strongly coupled. Finally, the total uncertainty, $\sigma_{\rm total}$, is obtained by quadratically summing up these errors and the line-to-line scatter ($\sigma_{\rm int}$). 

\begin{table*}
\caption{Mean NLTE abundances (the number of lines used for each element is given in brackets) with the usual notation assuming $\log \epsilon$(H)=12, and details on the error budget. The last column gives the total abundance uncertainty.}
\begin{center}
\begin{tabular}{lrc|ccccccc} \hline \hline
     &\multicolumn {2}{c}{Mean}    & \multicolumn {6}{c} {Error estimation} \\ 
     &\multicolumn {2}{c|} {abundance}  & $\sigma_{\rm int}  $   &  $\sigma_{\rm Teff}$  & $\sigma_{\rm logg}$ &  $\sigma_{\rm \xi}$ & $\sigma_{\rm Teff/logg}$ & $\sigma_{\rm total} $ \\ \hline
$\log \epsilon$(He) & 10.92& (8)  &  0.24 &  0.04 &  0.02 &  0.05 &  0.03 & 0.25 \\
$\log \epsilon$(C)  &  8.26& (9)  &  0.05 &  0.00 &  0.00 &  0.01 &  0.01 & 0.05 \\
$\log \epsilon$(N)  &  7.62& (22) &  0.08 &  0.07 &  0.03 &  0.03 &  0.03 & 0.12 \\
$\log \epsilon$(O)  &  8.43& (23) &  0.05 &  0.15 &  0.06 &  0.06 &  0.09 & 0.20 \\
$\log \epsilon$(Ne) &  8.26& (8)  &  0.05 &  0.05 &  0.01 &  0.01 &  0.04 & 0.08 \\
$\log \epsilon$(Mg) &  7.60& (1)  &   --- &  0.06 &  0.01 &  0.16 &  0.06 & 0.18 \\
$\log \epsilon$(Al) &  6.30& (3)  &  0.00 &  0.04 &  0.00 &  0.09 &  0.02 & 0.10 \\
$\log \epsilon$(Si) &  7.14& (8)  &  0.18 &  0.06 &  0.02 &  0.09 &  0.03 & 0.21 \\
$\log \epsilon$(Fe) &  7.30& (24) &  0.09 &  0.08 &  0.05 &  0.07 &  0.03 & 0.15 \\

\hline
\end{tabular}
\label{tab:abundances}
\end{center}
\end{table*}

 After completion of the analysis, we became aware of a study (Sim\'on-D\'{\i}az 2010) reporting problems with the modelling of some silicon lines using the model atom implemented in the NLTE code FASTWIND (Puls et al. 2005). Because our model atom is similar in many respects (Morel et al. 2006), we have examined to what extent this would affect our determination of the atmospheric parameters and ultimately abundances by repeating the abundance analysis excluding the spectral lines that may not be properly modelled (namely Si II $\lambda$5056 and Si III $\lambda$4813, 4819, 4829). This leads to a slightly lower temperature (see Fig.\,\ref{fig:Teff_logg}) and surface gravity: $T_{\rm eff}$ = 22300 K and $\log g$ = 3.72 (the microturbulence remains unchanged). The abundances determined using these parameters are provided in Table \ref{tab:alternative_abundances}. The differences with the mean abundances derived assuming the default parameters (Table \ref{tab:abundances}) remain within 0.1 dex, except in the case of Si where it amounts to about 0.18 dex.

\begin{table*}
\caption{Mean NLTE abundances (the number of lines used for each element is given in brackets) assuming $T_{\rm eff}$ = 22300 K, $\log g$ = 3.72 and $\xi$ = 1 km s$^{-1}$.}
\begin{center}
\begin{tabular}{lrc} \hline \hline    
     & \multicolumn {2}{c}{Mean}   \\
     & \multicolumn {2}{c}{abundance}  \\ \hline
$\log \epsilon$(He) & 10.88 & (8)  \\
$\log \epsilon$(C)  & 8.25  & (9)  \\
$\log \epsilon$(N)  & 7.65  & (22) \\
$\log \epsilon$(O)  & 8.52  & (23) \\
$\log \epsilon$(Ne) & 8.24  & (8)  \\
$\log \epsilon$(Mg) & 7.56  & (1)  \\
$\log \epsilon$(Al) & 6.32  & (3)  \\
$\log \epsilon$(Si) & 7.32  & (4)  \\
$\log \epsilon$(Fe) & 7.30  & (24) \\
\hline
\end{tabular}
\label{tab:alternative_abundances} 
\end{center}
\end{table*}

\section*{Discussion}
\subsection*{Chemical composition}
Figure \ref{fig:comparison_abundances} shows a comparison between our abundances (Table \ref{tab:abundances}) and previous NLTE results in the literature (Gies \& Lambert 1992, Andrievsky et al. 1999, Korotin et al. 1999a,b,  Morel et al. 2006, Morel \& Butler 2008), as well as the new solar abundances of Asplund et al. (2009). There is an overall satisfactory agreement with these previous studies. However, all these results taken at face value would suggest that $\gamma$ Peg is metal poor with respect to the Sun, a fact which is obviously contrary to the expectations for a young star in the solar neighbourhood. This is likely connected to the long-standing and more general problem affecting most abundance analyses of OB stars, which yield abundances significantly lower than solar (e.g., Morel 2009 for a review). Although the origin of this problem remains unclear, improvements in the atomic data and determination of the atmospheric parameters have been claimed to solve most of the discrepancy (Przybilla et al. 2008, Sim\'on-D\'{\i}az 2010). In view of this rather unsatisfactory situation, a sound assumption may be to use in further theoretical modelling the abundances determined for a small sample of nearby B-type stars by Przybilla et al. (2008) or the solar abundances of Asplund et al. (2009). 

\figureDSSN{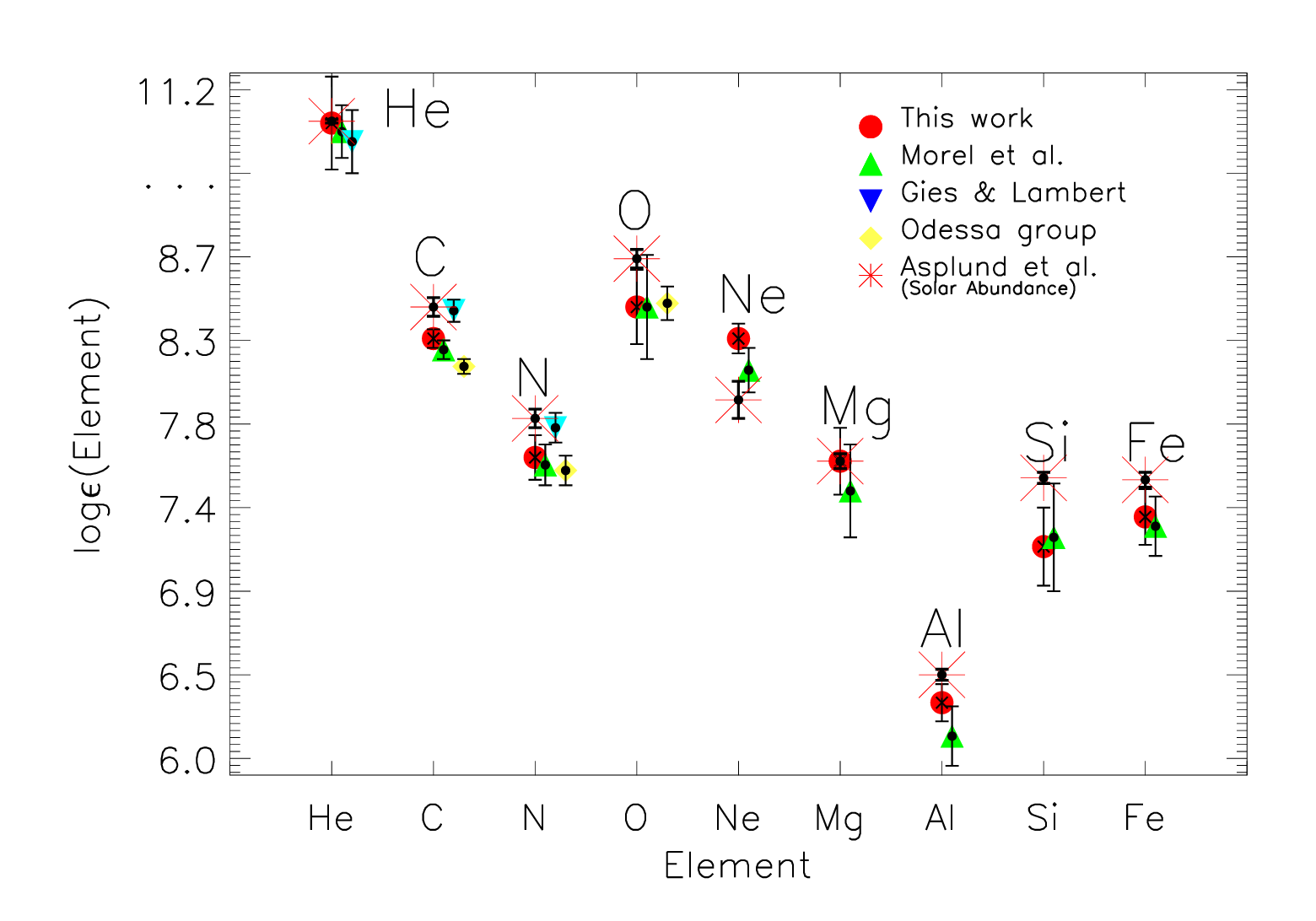}{Comparison between our abundances and previous NLTE results in the literature (Gies \& Lambert 1992, Morel et al. 2006, Morel \& Butler 2008, Andrievsky et al. 1999, Korotin et al. 1999a,b). The data points dubbed 'Odessa group' refer to the results of Andrievsky et al. (1999) and Korotin et al. (1999a,b). The solar abundances of Asplund et al. (2009) are indicated with different symbols.}{fig:comparison_abundances}{!ht}{clip,trim=1.3cm 0cm 1cm 1cm, angle=0,width=1\textwidth}

However, a robust result is that the chemical composition of $\gamma$ Peg does not significantly differ from that of early B-type stars analysed using similar techniques, whether they are known as pulsating or not (Morel et al. 2008). The existence of microscopic diffusion processes in hybrid pulsators leading to an accumulation of iron in the driving zone has often been invoked to explain the unexpected excitation of some specific pulsation modes (e.g., Pamyatnykh et al. 2004), but the lack of any abundance peculiarities does not seem at first glance to support this possibility unless one supposes that the changes at the surface remain below the detection limits. On the other hand, the [N/O] and [N/C] logarithmic abundance ratios (--0.81$\pm$0.24 and --0.64$\pm$0.13 dex, respectively) are identical to within the errors to the solar values (Asplund et al. 2009). Our study therefore confirms that there is no evidence in $\gamma$ Peg for CNO-processed material dredged up to the surface because of deep mixing. The significant boron depletion (Proffitt \& Quigley 2001) suggests, however, that shallow mixing is already taking place in the superficial layers. 

\subsection*{Comparison with theoretical instability strips}
As $\gamma$ Peg is one of the rare stars known to date to present both high-order $g$ modes and low-order $p$ and $g$ modes, it is of interest to examine whether this property is consistent with the theoretical expectations. The position of $\gamma$ Peg in the ($\log T_{\rm eff}$, $\log g$) plane is shown in Fig.\,\ref{fig:instability_strips}, along with the instability domains for $\beta$ Cephei and SPB-like pulsation modes computed with the solar abundances of Asplund et al. (2005) and for two different opacity tables, OP and OPAL (see Miglio et al. 2007 for details). At face value, this figure suggests that the hybrid nature of the pulsations in $\gamma$ Peg is more consistent with instability strips computed with OP opacities, as the OPAL ones do not predict this star to exibit SPB-like pulsations. 

\figureDSSN{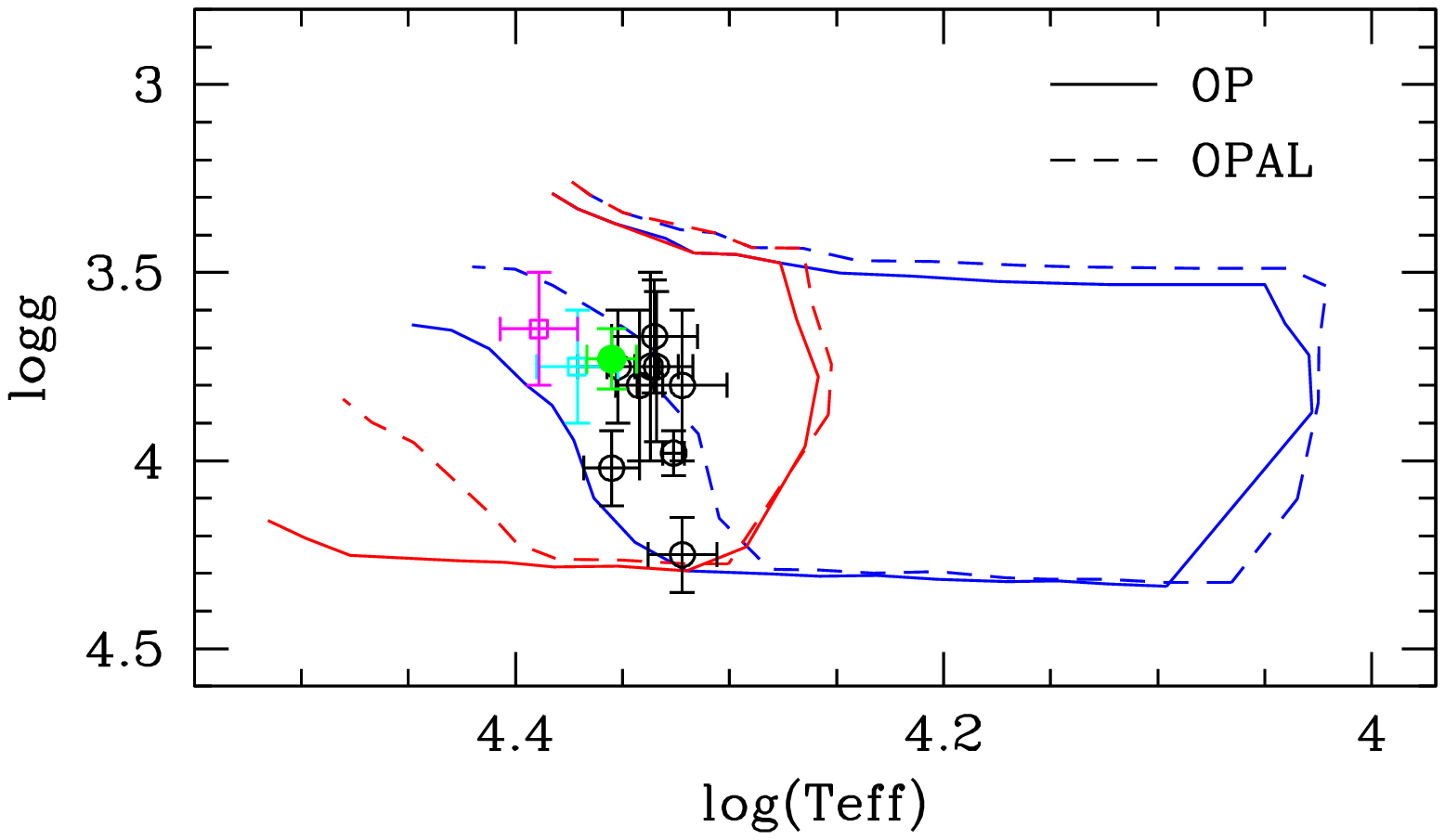}{Position of $\gamma$ Peg in the ($\log T_{\rm eff}$, $\log g$) plane (green filled circle), along with the instability strips for $\beta$ Cephei- (red lines) and SPB-like (blue lines) pulsation modes ($l$ $<$ 4) computed using the solar abundances of Asplund et al. (2005) and the OP and OPAL opacity tables (Miglio et al. 2007). The neon abundance is taken from Cunha et al. (2006). The open circles indicate results for $\gamma$ Peg in the literature (Fitzpatrick \& Massa 2005, Gies \& Lambert 1992, Korotin et al. 1999a, Martin 2004, Morel et al. 2006 [based on Si ionisation equilibrium], Morel \& Butler 2008 [based on Ne ionisation equilibrium], Niemczura \& Po\l ubek 2006, Pintado \& Adelman 1993, Ryans et al. 1996). The two open squares refer to the results for two other hybrid pulsators, $\nu$ Eri and 12 Lac (Morel et al. 2006).}{fig:instability_strips}{!ht}{clip,angle=0,width=.9\textwidth}

As can also be seen in this figure, the same conclusion holds for the two other best-studied hybrid pulsators, $\nu$ Eri (Jerzykiewicz et al. 2005) and 12 Lac (Handler et al. 2006), that have been analysed using similar techniques as in the present work (Morel et al. 2006). Identical parameters within the errors were obtained for $\nu$ Eri by De Ridder et al. (2004) from a full spectroscopic analysis. Lefever et al. (2010) analysed $\gamma$ Peg, $\nu$ Eri and 12 Lac using the same data as Morel et al. (2006), but with a different code, and also found very similar results, except a temperature lower by 1500 K in 12 Lac. These two studies also determined $T_{\rm eff}$ and $\log g$ from Si ionisation balance and fitting of the Balmer lines. On the other hand, and although they may be regarded as less reliable as they heavily rely on only one very weak Ne II line, slightly lower $T_{\rm eff}$ values were obtained for these three stars by Morel \& Butler (2008) based on Ne ionisation equilibrium. Finally, Gies \& Lambert (1992) found significantly higher $T_{\rm eff}$ values for $\nu$ Eri and 12 Lac based on photometric data, but it is likely that their temperature scale is too hot (see discussion in Lyubimkov et al. 2002). This also leads to surface gravities that are higher than other results in the literature.  

While the opacity values prove critical for the excitation, or lack thereof, of oscillations in B stars, the location of the SPB instability domain is relatively insensitive to the chemical mixture used (Miglio et al. 2007). Using the recently revised abundances of Asplund et al. (2009) is therefore not expected to change this picture significantly (an identical situation is indeed encountered when using the more different solar mixture of Grevesse \& Noels 1993). 

To investigate to what extent our conclusions regarding the ability of the opacity calculations to reproduce the observations is robust against the choice of $\log T_{\rm eff}$ and $\log g$, we show in Fig.\,\ref{fig:instability_strips} other results for $\gamma$ Peg from the literature. Although a variety of techniques have been used for the determination of the parameters, almost all these results rely on photometric data and/or LTE methods. It should also be noted that these estimates are not completely independent, as the same photometric calibrations (and data) have often been employed. Despite these caveats, the gravity seems well constrained to around $\log g$ $\sim$ 3.8 if one excludes the discrepant value of Pintado \& Adelman (1993) based on LTE fitting of the H$\gamma$ line. The effective temperatures are generally based on narrow-band photometric data and appear slightly lower than our value based on Si ionisation equilibrium (keeping in mind that, as discussed above, rejecting some potentially problematic Si lines would lower our $T_{\rm eff}$ by about 350 K). Taking into account the still relatively large inaccuracy of the $T_{\rm eff}$ value, we conclude, solely based on the position of $\gamma$ Peg with respect to the instability strips, that the OP opacities are only marginally preferred to the OPAL ones. Accurate positions for a sufficiently large sample of stars may lead to stronger statements. It should also be kept in mind, however, that some important physical processes that are not currently implemented in the stability analysis may substantially change the results of the models (e.g., stellar rotation; Townsend 2005). 

To conclude, although it is clear that several theoretical aspects related to the excitation/damping of oscillations in B stars may greatly benefit from the constraints offered by the further identification of hybrid pulsators and from a detailed study of their pulsation properties, an accurate determination of their fundamental parameters remains essential for further progress.

\acknowledgments{ \small We are indebted to Dr. K. Butler for his valuable comments and for kindly providing us with the codes DETAIL/SURFACE. C.P.P. sincerely thanks to the members of the technical staff (M. Appakutti, V. Moorthy and C. Velu) at VBO  (Kavalur, India) for their kind assistance during the observations, Profs. N. K. Rao, S. Giridhar, H. C. Bhatt, and B. P. Das, Drs. C. Muthumariappan and  G. Pandey for their kind hospitality and encouragement during her visits to the Indian Institute of Astrophysics (Bengaluru, India), as well as Dr. U. S. Chaubey for his encouragement and for providing the financial assistance for her visits to VBO, through a sponsored project (Grant No SR/S2/HEP-20/2003) of the Department of Science and Technology (DST), Government of India, New Delhi and also Prof. Ram Sagar for his kind motivation. T. M. acknowledges financial support from Belspo for contract PRODEX-GAIA DPAC. M.B. is Postdoctoral Fellow of the Fund for Scientific Research, Flanders. We are grateful to Andrea Miglio for kindly providing us the theoretical instability strips for B-type pulsators and for useful comments. The authors are also grateful to the anonymous referee for his/her constructive comments which helped in improving the presentation of the paper. This research made use of NASA Astrophysics Data System Bibliographic Services, the SIMBAD database operated at CDS, Strasbourg (France). }

\References{

\rfr Aerts, C., de Pauw, M., \& Waelkens, C. 1992, A\&A, 266, 294

\rfr Aller, L. H. 1949, ApJ, 109, 244

\rfr Andrievsky, S. M.,  Korotin, S. A., Luck, R. E., \& Kostynchuk, L. Yu. 1999, A\&A, 350, 598 

\rfr Asplund, M., Grevesse, N., \& Sauval, A. J. 2005, ASP Conf. Ser., 336, 25

\rfr Asplund, M., Grevesse, N., Sauval, A. J., \& Scatt, P. 2009, ARA\&A 47, 481

\rfr Balona, L. A., Pigulski, A., De Cat, P., et al. 2011, MNRAS, in press

\rfr Breger, M., Stich, J., Garrido, R., et al. 1993, A\&A, 271, 482

\rfr Briquet, M., Lefever, K., Uytterhoeven, K., \& Aerts, C. 2005, MNRAS, 362, 619

\rfr Butkovskaya, V. V., \& Plachinda, S. I. 2007, A\&A, 469, 1069

\rfr Butler, K., \& Giddings, J. R. 1985, Newsletter of Analysis of Astronomical Spectra, No.9 (Univ. London)

\rfr Castelli, F. 1991, A\&A, 251, 106

\rfr Castelli, F., \& Kurucz, R. L. 2004, IAU Symp., 210, A20 (arXiv:astro-ph/0405087)

\rfr Chapellier, E., Le Contel, D.,  Le Contel, J. M., et al. 2006, A\&A, 448, 697

\rfr Crawford, D. L., Barnes, J. V., \& Golson, J.C. 1971, AJ, 76, 1058

\rfr Cunha, K., Hubeny, I., \& Lanz, T. 2006, ApJ,  647, 143

\rfr De Cat, P., Briquet, M., Aerts, C., et al. 2007, A\&A, 463, 243

\rfr Degroote, P.,  Briquet, M., Catala, C., et al.  2009, A\&A, 506, 111   

\rfr De Ridder, J., Telting, J. H., Balona, L. A., et al. 2004, MNRAS, 351, 324

\rfr Desmet, M., Briquet, M., Thoul, A., et al. 2009, MNRAS, 396, 1460

\rfr Fitzpatrick, E. L., \& Massa, D. 2005, AJ, 129, 1642

\rfr Giddings, J. R. 1981, Ph.D. Thesis , University of London

\rfr Gies, D. R., \& Lambert, D. L. 1992, ApJ, 387, 673

\rfr Grevesse, N., \& Noels, A. 1993, in Hauck B. and Paltani S. R. D., eds, La Formation des El\'ements Chimiques. AVCP, Lausanne, 205

\rfr Grevesse, N., \& Sauval, A. J., 1998, Space Sci. Rev., 85, 161

\rfr Grigahc\`ene, A.,  Antoci, V., Balona, L., et al. 2010, ApJ, 713, L192

\rfr Handler, G., Jerzykiewicz, M., Rodriguez, E., et al. 2006, MNRAS, 365, 327 

\rfr Handler, G. 2009, MNRAS, 398, 1339 

\rfr Handler, G., Matthews, J. M., Eaton, J. A., et al. 2009, ApJ, 698, 56 

\rfr Harmanec, P., Koubsky, P., Krpata, J., \& Zdarsky, F. 1979, IBVS, 1590

\rfr Hauck, B., \&  Mermilliod, M. 1998, A\&AS,  129, 431

\rfr Jerzykiewicz, M., Handler, G., Shobbrook, R. R., et al. 2005,  MNRAS, 360, 619

\rfr Korotin, S. A., Andrievsky, S.M., \& Kostynchuk, L. Yu. 1999a, A\&A, 342, 756

\rfr Korotin, S. A., Andrievsky, S. M., \& Luck, R. E. 1999b, A\&A, 351, 168

\rfr Kurucz, R. L. 1993, ATLAS9 Stellar Atmosphere Programs and 2 km/s grid. Kurucz CD-ROM No. 13. Cambridge, Mass.: Smithsonian Astrophysical Observatory, 13

\rfr Lefever, K., Puls, J., Morel, T., et al. 2010,  A\&A , 515, A74

\rfr Lyubimkov, L. S., Rachkovkaya, T. M., Rostopchin, S. I., et al. 2002, MNRAS, 333, 9

\rfr Martin, J. C. 2004, AJ, 128, 2474

\rfr Mazumdar, A., Briquet, M., Desmet, M., \& Aerts, C. 2006, A\&A, 459, 589

\rfr Miglio, A., Montalb\'an, J., \& Dupret, M.-A. 2007, MNRAS, 375, 21

\rfr Morel, T., Butler, K., Aerts, C., et al. 2006, A\&A, 457, 651

\rfr Morel, T., Hubrig, S., \& Briquet, M. 2008, A\&A, 481, 453

\rfr Morel, T., \& Butler, K. 2008, A\&A, 487, 307 

\rfr Morel, T. 2009, CoAst, 158, 122 

\rfr Niemczura, E., \& Po\l ubek, G. 2006, ESASP, 624, 120

\rfr Nieva, M. -F., \& Przybilla, N. 2007, A\&A, 467, 295

\rfr Pamyatnykh, A. A., Handler, G., \& Dziembowski, W. A. 2004,  MNRAS, 350, 1022

\rfr Pigulski, A., \& Pojma\'nski, G. 2008, A\&A, 477, 917 

\rfr Pintado, O. I., \& Adelman, S. J. 1993, MNRAS, 264, 63

\rfr Press, W. H., Teukolsky, S. A., Vetterling, W. T., \& Flannery, B. P.,  Numerical Recipes 3rd Edition: The Art of Scientific Computing, Cambridge University Press, New York, 2007

\rfr Proffitt, C. R., \& Quigley, M. F. 2001,  ApJ, 548, 429

\rfr Przybilla, N., Nieva, M. F., \& Butler, K. 2008, ApJ, 688, 103 

\rfr Puls, J., Urbaneja, M. A., Venero, R., et al. 2005, A\&A, 435, 669

\rfr Rao, N. K., Sriram, S., Jayakumar, K., \& Gabriel, F. 2005, JAA, 26, 331

\rfr Ryans, R. S. I., Hambly, N. C., Dufton, P. L., \&  Keenan, F. P. 1996, MNRAS, 278, 132

\rfr Sim\'on-D\'{\i}az, S. 2010, A\&A, 510, 22

\rfr Sterken, C., \& Jerzykiewicz, M. 1993, SSRv, 62,95

\rfr Thoul, A. 2009, CoAst, 159,35

\rfr Townsend, R. H. D. 2005, MNRAS, 360, 465 

\rfr Walczak, P., \& Daszy\'nska-Daszkiewicz, J. 2010, AN, 331, 1057

\rfr Zdravkov, T., \& Pamyatnykh, A. A. 2009, AIPC, 1170, 388 

\rfr Zima, W. 2006, A\&A, 455, 227

\rfr Zima, W. 2008, CoAst, 155, 1

}

\newpage

\appendix
\section{Appendix}

\renewcommand{\thefigure}{\arabic{figure}\alph{subfigure}}

\addtocounter{figure}{0}
\setcounter{subfigure}{1}
\figureDSSN{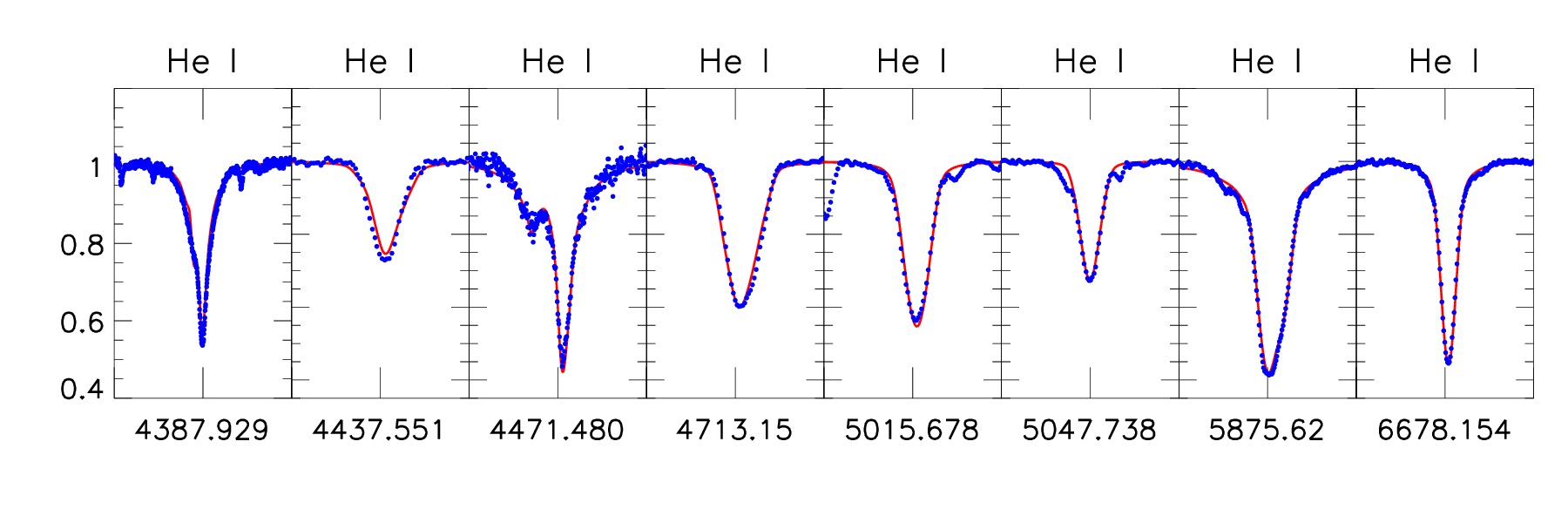}{Comparison between the observed and synthetic profiles (represented by the dots and full lines, respectively) computed for the abundance yielded by the corresponding He I line. The synthetic spectra have been convolved with a rotational broadening function with $v_T$=8 km s$^{-1}$.}{fig:lineprofilesHe}{!ht}{clip,angle=0,width=1\textwidth}

\vskip 1cm

\addtocounter{figure}{-1}
\addtocounter{subfigure}{1}
\figureDSSN{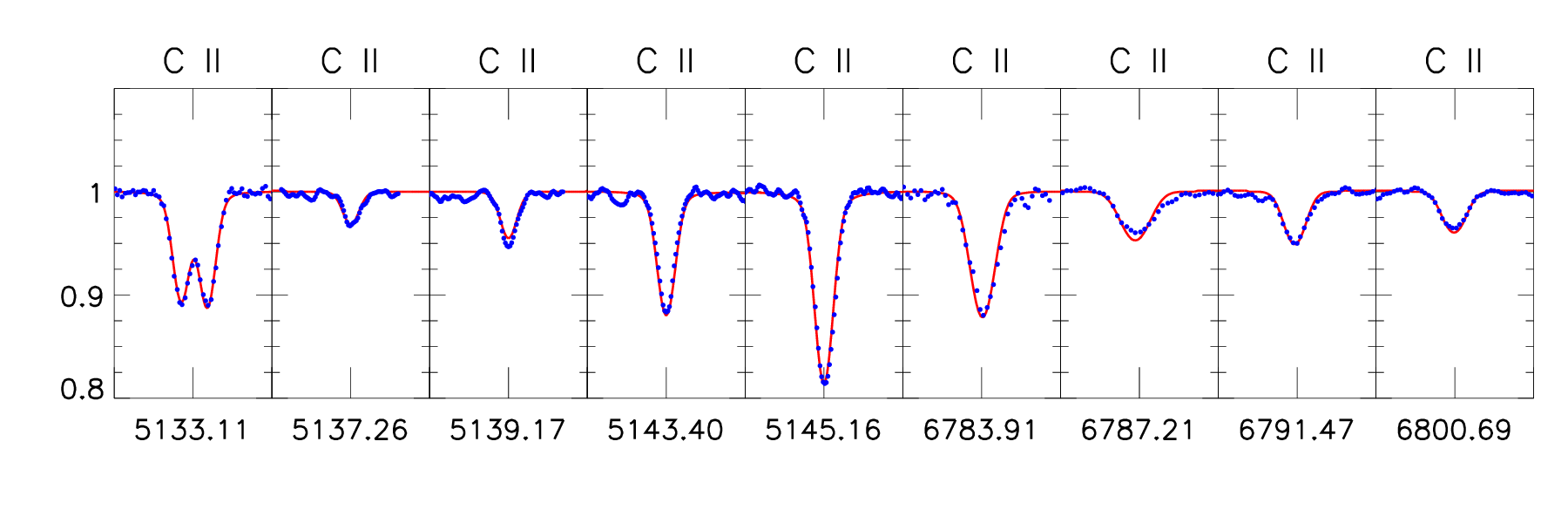}{Same as Fig.\,\ref{fig:lineprofilesHe},  but for the C II lines.}{fig:lineprofilesC}{!ht}{clip, angle=0,width=\textwidth}

\addtocounter{figure}{-1}
\addtocounter{subfigure}{1}
\figureDSSN{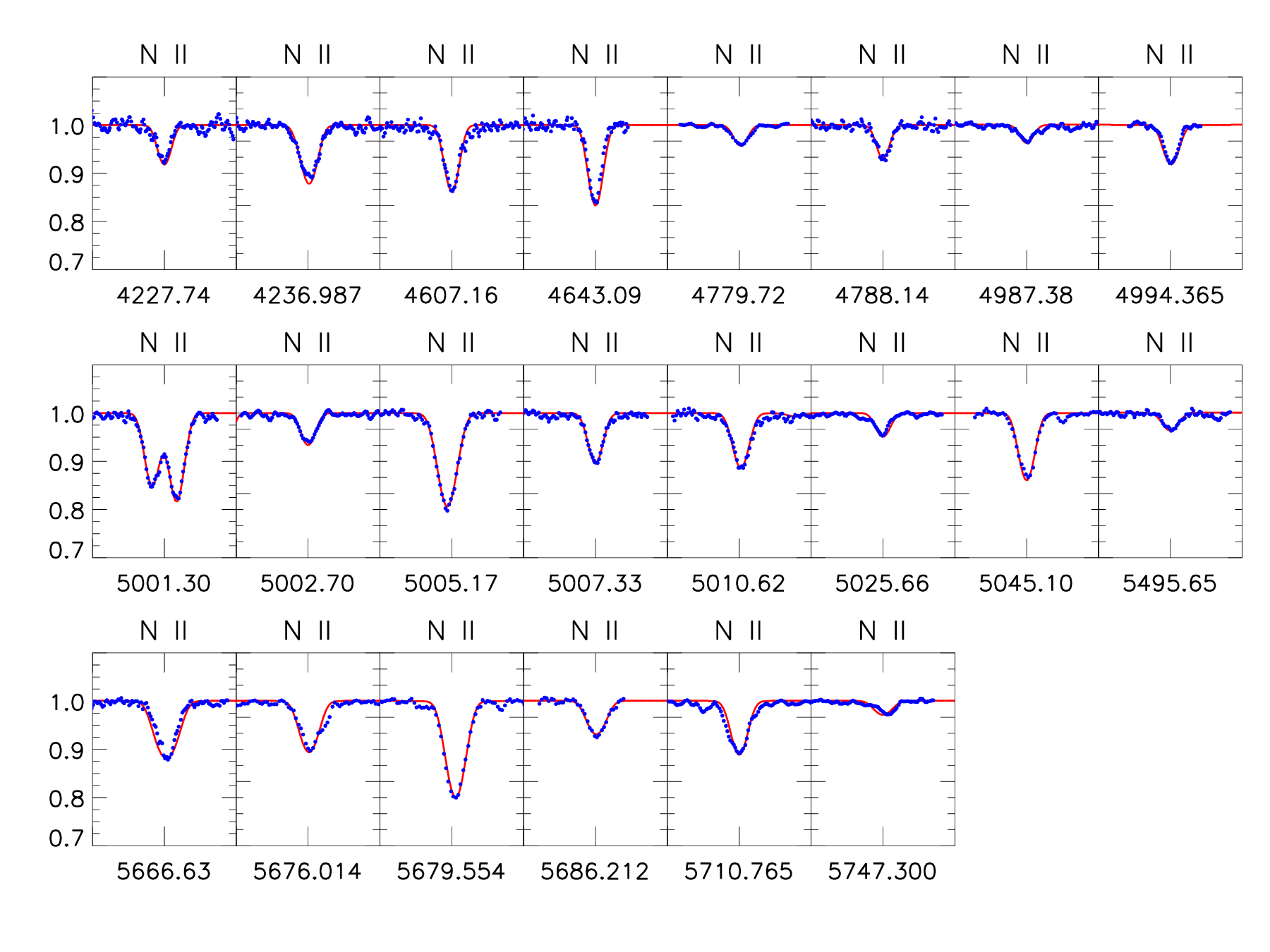}{Same as Fig.\,\ref{fig:lineprofilesHe},  but for the N II lines.}{fig:lineprofilesN}{!ht}{clip,angle=0, width=\textwidth}

\addtocounter{figure}{-1}
\addtocounter{subfigure}{1}
\figureDSSN{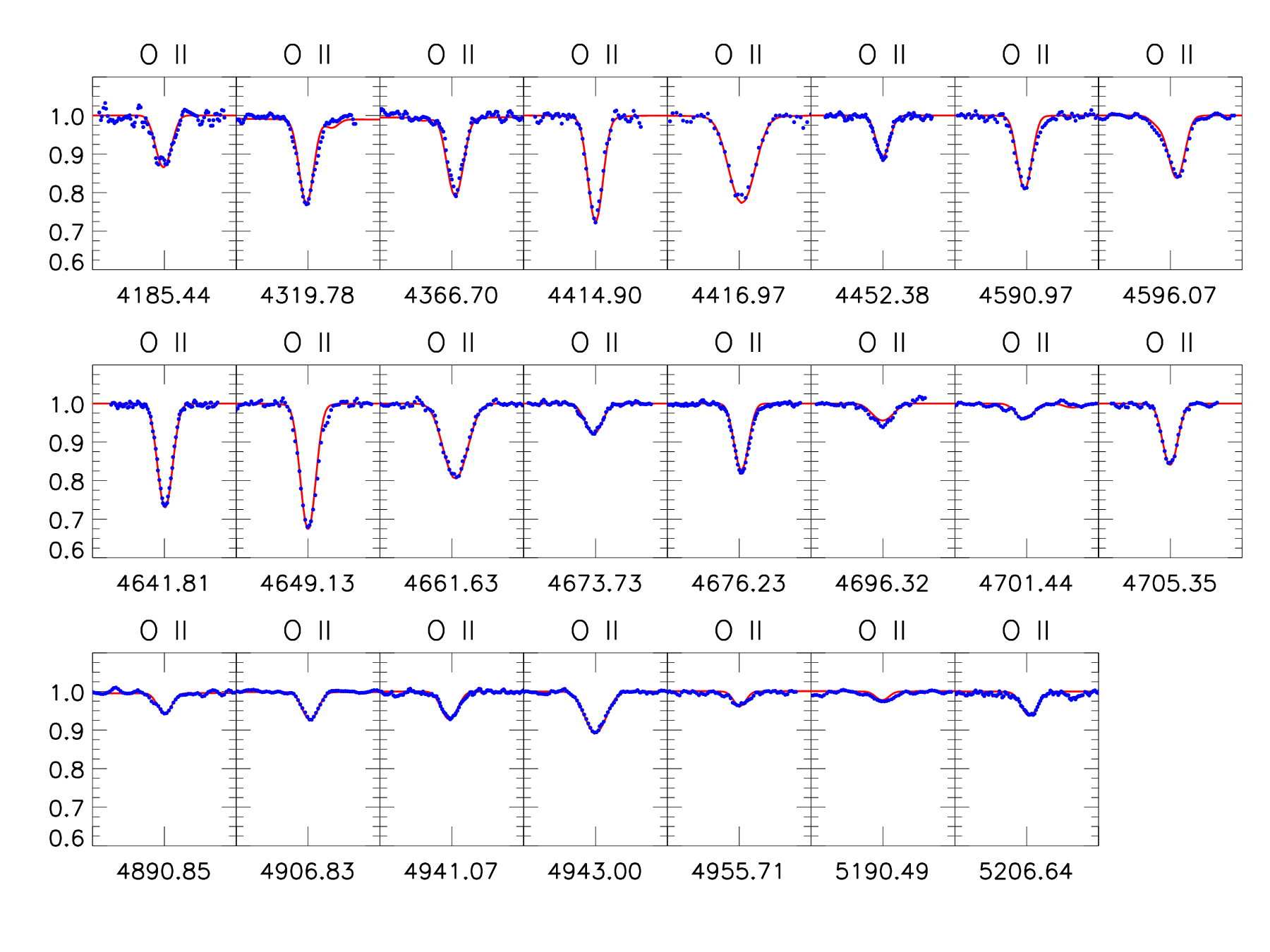}{Same as Fig.\,\ref{fig:lineprofilesHe},  but for the O II lines.}{fig:lineprofilesO}{!ht}{clip,angle=0, width=\textwidth}

\addtocounter{figure}{-1}
\addtocounter{subfigure}{1}
\figureDSSN{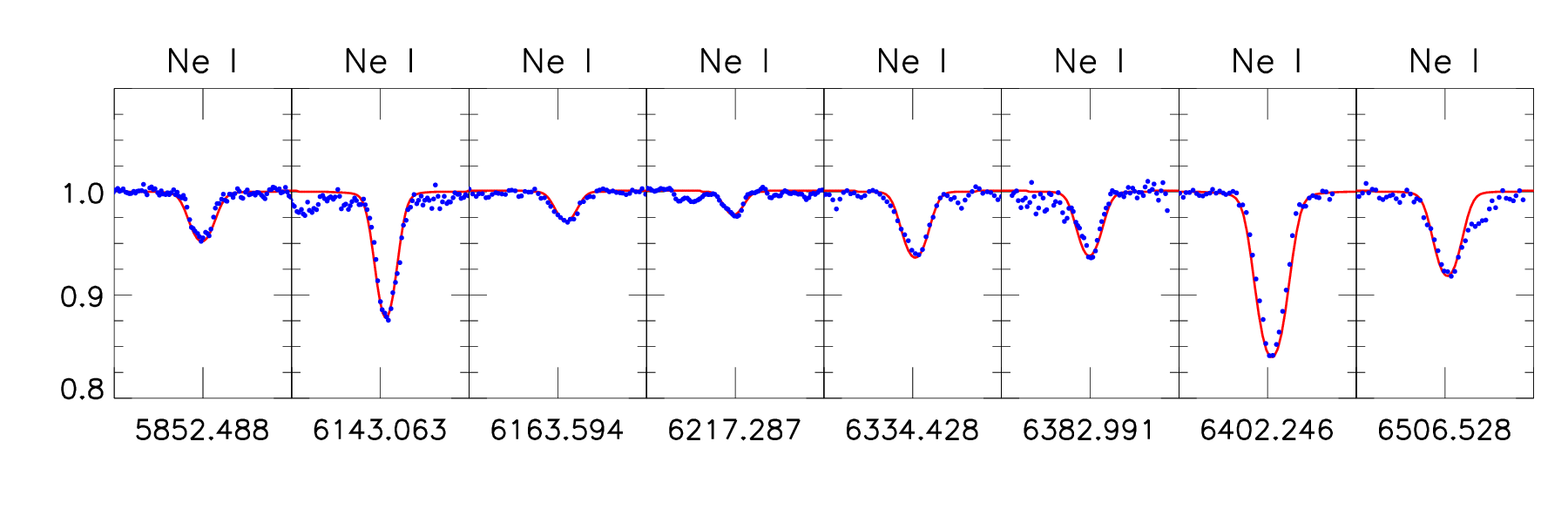}{Same as Fig.\,\ref{fig:lineprofilesHe}, but for the Ne I lines.}{fig:lineprofilesNe}{!ht}{clip,angle=0,width=\textwidth}

\addtocounter{figure}{-1}
\addtocounter{subfigure}{1}
\figureDSSN{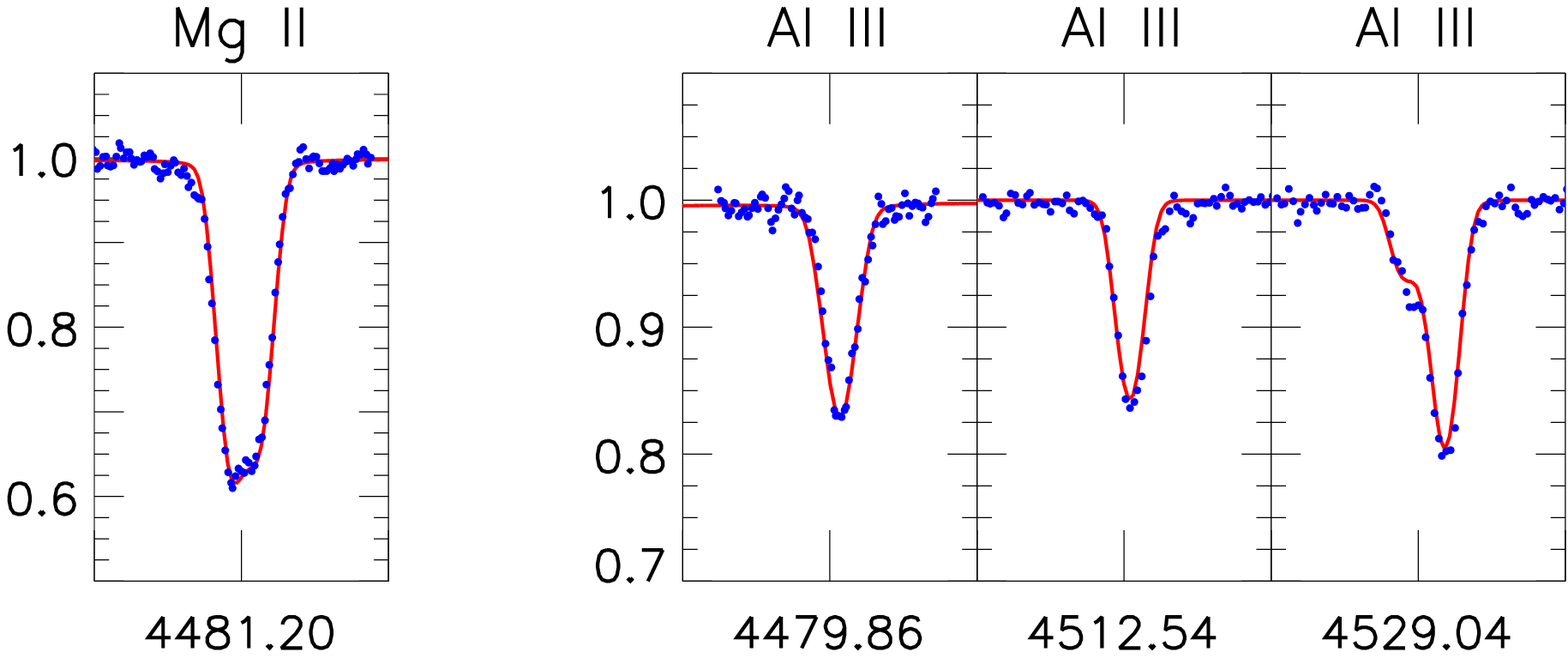}{Same as Fig.\,\ref{fig:lineprofilesHe}, but for the Mg II (left) and Al III (right) lines.}{fig:lineprofilesMgAl}{!ht}{clip,angle=0,height=.16\textheight,width=.8\textwidth}

\addtocounter{figure}{-1}
\addtocounter{subfigure}{1}
\figureDSSN{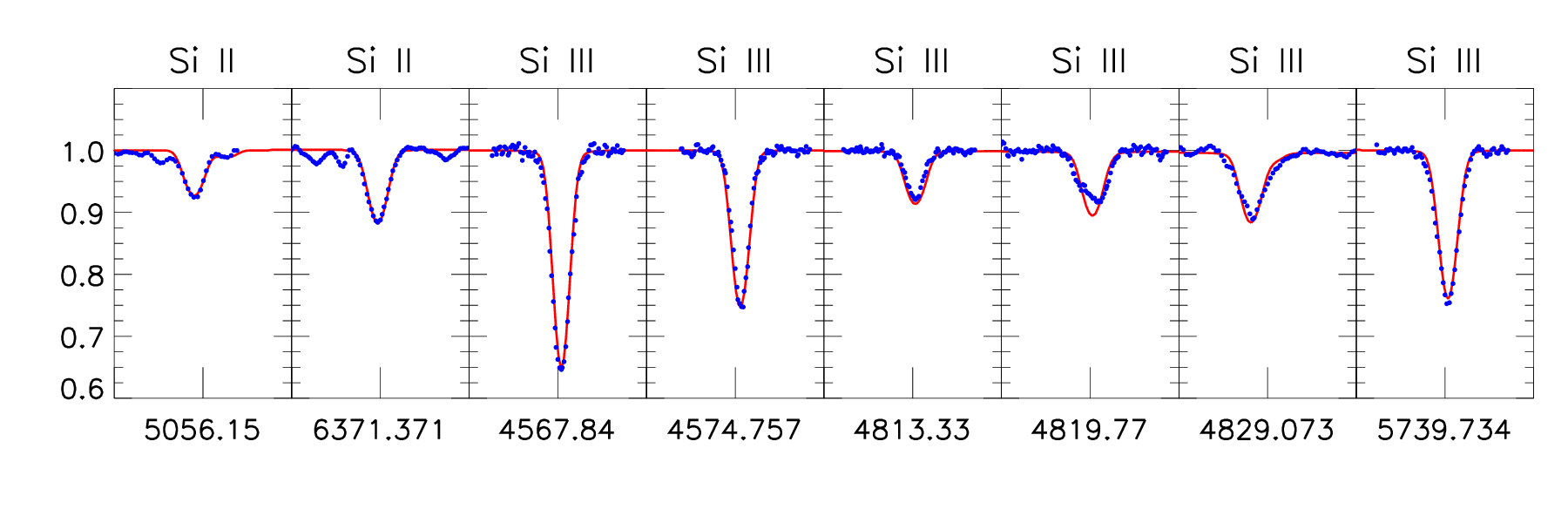}{Same as Fig.\,\ref{fig:lineprofilesHe}, but for the Si II and Si III lines.}{fig:lineprofilesSi}{!ht}{clip,angle=0,width=\textwidth}

\addtocounter{figure}{-1}
\addtocounter{subfigure}{1}
\figureDSSN{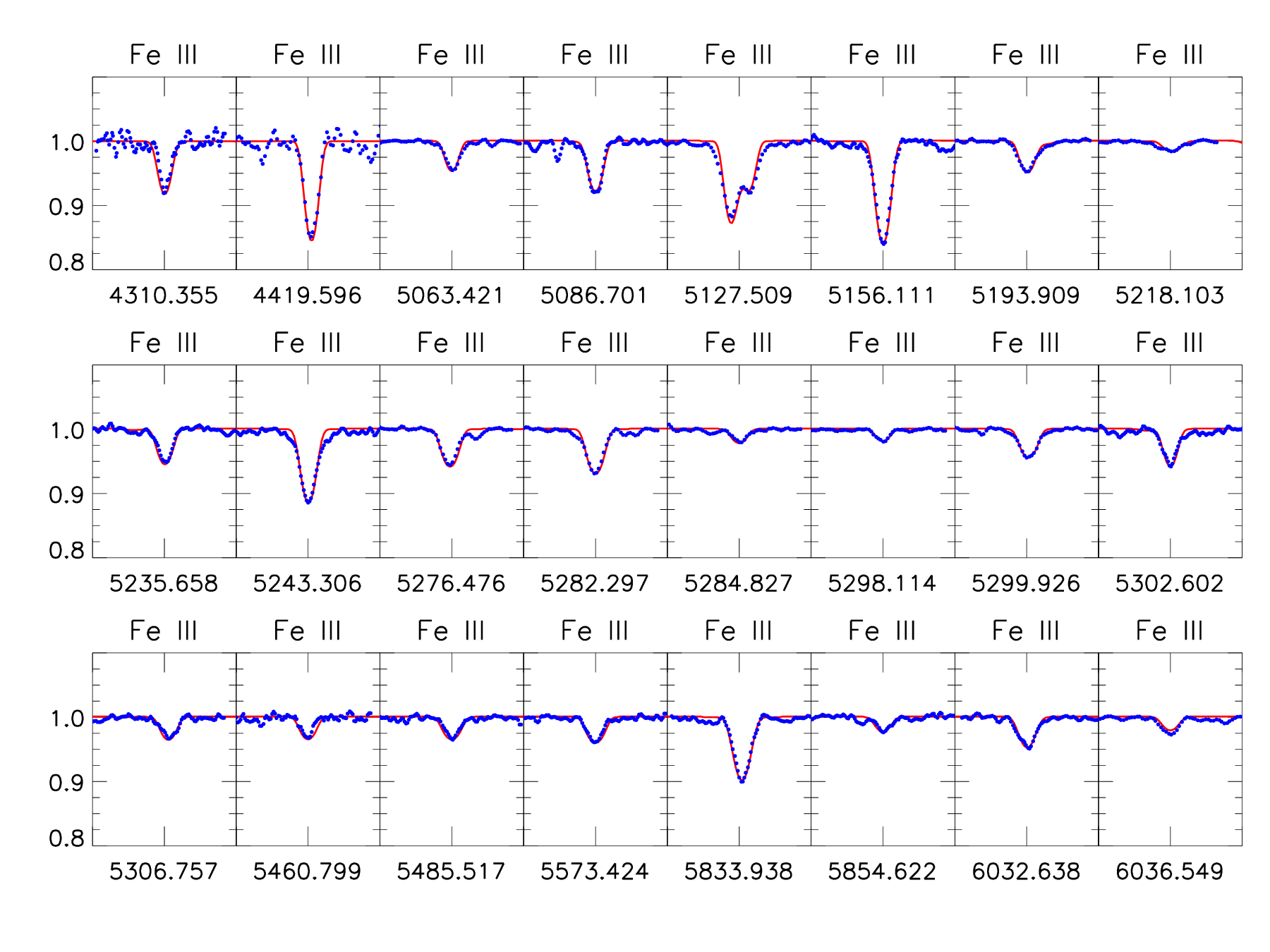}{Same as Fig.\,\ref{fig:lineprofilesHe}, but for the Fe III lines.}{fig:lineprofilesFe}{!ht}{clip,angle=0, width=\textwidth}

\newpage

\begin{table*}[!ht]
\caption{EW measurements and abundance results for each transition, along with the error estimates.}
\label{tab:individual_abundances}
\begin{center}
\begin{tabular}{lccc|ccccc} \hline\hline
{Ion}     &{ Transition }& {EW}& $\log \epsilon$ & \multicolumn {5}{c} {Error estimation} \\ 
 & (\AA)   & (m\AA)  & (dex)   & {$\sigma_{\rm Teff}$ } & {$\sigma_{\rm logg} $} & {$\sigma_ {\rm \xi} $} & {$\sigma_{\rm Teff/logg}$ } & {$\sigma_{\rm total} $} \\ \hline
{\bf He I}

          & 4387.93 & 820  & 11.29 & 0.08 & 0.03 & 0.00 & 0.04 & 0.09 \\ 
          & 4437.55 & 109  & 10.98 & 0.06 & 0.00 & 0.06 & 0.06 & 0.10 \\ 
          & 4471.48 & 1255 & 11.09 & 0.05 & 0.04 & 0.00 & 0.00 & 0.06 \\ 
          & 4713.15 & 270  & 10.92 & 0.07 & 0.00 & 0.07 & 0.07 & 0.12 \\ 
          & 5015.68 & 274  & 10.76 & 0.00 & 0.09 & 0.10 & 0.00 & 0.13 \\ 
          & 5047.74 & 170  & 11.04 & 0.06 & 0.00 & 0.06 & 0.06 & 0.10 \\ 
          & 5875.62 & 605  & 10.53 & 0.00 & 0.00 & 0.00 & 0.00 & 0.00 \\ 
          & 6678.15 & 560  & 10.76 & 0.00 & 0.00 & 0.10 & 0.00 & 0.10 \\ 
          &&&&&&&&\\

{\bf C II}  

         & 5133.11  & 62 & 8.22 & 0.01 & 0.00 & 0.01 & 0.01 & 0.02 \\ 
         & 5137.26  &  9 & 8.31 & 0.00 & 0.00 & 0.00 & 0.01 & 0.01 \\
         & 5139.17  & 14 & 8.33 & 0.01 & 0.01 & 0.01 & 0.02 & 0.03 \\
         & 5143.40  & 33 & 8.26 & 0.01 & 0.00 & 0.01 & 0.01 & 0.02 \\
         & 5145.16  & 58 & 8.25 & 0.01 & 0.01 & 0.03 & 0.01 & 0.03 \\
         & 6783.91  & 50 & 8.25 & 0.00 & 0.01 & 0.03 & 0.01 & 0.03 \\
         & 6787.21  & 17 & 8.30 & 0.01 & 0.00 & 0.01 & 0.01 & 0.02 \\
         & 6791.47  & 18 & 8.23 & 0.00 & 0.01 & 0.01 & 0.01 & 0.02 \\
         & 6800.69  & 15 & 8.20 & 0.01 & 0.00 & 0.01 & 0.01 & 0.02 \\
         &&&&&&&\\

{\bf N II} 

         & 4227.74 & 16  & 7.53  &  0.05 &   0.01 &   0.02 &  0.03 &  0.06  \\ 
         & 4236.99 & 38  & 7.48  &  0.06 &   0.02 &   0.01 &  0.03 &  0.07  \\
         & 4607.16 & 40  & 7.73  &  0.07 &   0.02 &   0.03 &  0.03 &  0.08  \\
         & 4643.09 & 42  & 7.62  &  0.06 &   0.03 &   0.04 &  0.03 &  0.08  \\
         & 4779.72 & 11  & 7.66  &  0.05 &   0.01 &   0.00 &  0.02 &  0.05  \\
         & 4788.14 & 18  & 7.70  &  0.06 &   0.02 &   0.02 &  0.03 &  0.07  \\
         & 4987.38 & 10  & 7.70  &  0.05 &   0.02 &   0.01 &  0.03 &  0.06  \\
         & 4994.36 & 19  & 7.53  &  0.07 &   0.02 &   0.02 &  0.04 &  0.09  \\
         & 5001.30 & 83  & 7.59  &  0.09 &   0.03 &   0.05 &  0.05 &  0.12  \\
         & 5002.70 & 16  & 7.72  &  0.05 &   0.03 &   0.02 &  0.02 &  0.06  \\
         & 5005.17 & 53  & 7.55  &  0.10 &   0.03 &   0.06 &  0.06 &  0.13  \\
         & 5007.33 & 28  & 7.56  &  0.08 &   0.02 &   0.03 &  0.04 &  0.10  \\
         & 5010.62 & 29  & 7.65  &  0.06 &   0.03 &   0.03 &  0.02 &  0.08  \\
         & 5025.66 & 13  & 7.76  &  0.07 &   0.02 &   0.01 &  0.04 &  0.08  \\
         & 5045.10 & 39  & 7.65  &  0.06 &   0.04 &   0.04 &  0.03 &  0.09  \\
         & 5495.65 & 10  & 7.53  &  0.07 &   0.02 &   0.01 &  0.04 &  0.08  \\
         & 5666.63 & 50  & 7.57  &  0.09 &   0.04 &   0.05 &  0.05 &  0.12  \\
         & 5676.01 & 35  & 7.64  &  0.08 &   0.04 &   0.04 &  0.04 &  0.11  \\
         & 5679.55 & 71  & 7.58  &  0.10 &   0.04 &   0.07 &  0.05 &  0.14  \\
         & 5686.21 & 21  & 7.50  &  0.07 &   0.03 &   0.02 &  0.03 &  0.08  \\
         & 5710.77 & 33  & 7.76  &  0.08 &   0.04 &   0.04 &  0.04 &  0.11  \\
         & 5747.30 &  8  & 7.59  &  0.06 &   0.03 &   0.01 &  0.03 &  0.07  \\
               
         \hline
\end{tabular}
\end{center}
\end{table*}

\newpage

\begin{table*}[h]
\begin{center}
\begin{tabular}{lccc|ccccc} 
\multicolumn {8}{l} {continued...}\\
\\
 {\bf O II} & 4185.44 &  30  &   8.42 &  0.15 &  0.05 &   0.07 &  0.10 &  0.20 \\ 
            & 4319.78 &  60  &   8.60 &  0.15 &  0.06 &   0.08 &  0.10 &  0.21 \\
            & 4366.70 &  48  &   8.40 &  0.15 &  0.06 &   0.08 &  0.09 &  0.20 \\
            & 4414.90 &  70  &   8.37 &  0.15 &  0.05 &   0.09 &  0.10 &  0.21 \\
            & 4416.97 &  56  &   8.41 &  0.15 &  0.06 &   0.09 &  0.09 &  0.21 \\
            & 4452.38 &  25  &   8.44 &  0.13 &  0.05 &   0.04 &  0.08 &  0.17 \\
            & 4590.97 &  52  &   8.48 &  0.18 &  0.06 &   0.08 &  0.12 &  0.24 \\
            & 4596.07 &  50  &   8.47 &  0.16 &  0.06 &   0.05 &  0.10 &  0.20 \\
            & 4641.81 &  68  &   8.40 &  0.18 &  0.06 &   0.10 &  0.11 &  0.24 \\
            & 4649.13 &  92  &   8.51 &  0.17 &  0.07 &   0.14 &  0.11 &  0.26 \\
            & 4661.63 &  47  &   8.37 &  0.16 &  0.06 &   0.08 &  0.10 &  0.21 \\
            & 4673.73 &  17  &   8.42 &  0.12 &  0.05 &   0.03 &  0.08 &  0.16 \\
            & 4676.23 &  45  &   8.46 &  0.16 &  0.06 &   0.07 &  0.10 &  0.21 \\
            & 4696.32 &  10  &   8.38 &  0.12 &  0.05 &   0.02 &  0.07 &  0.15 \\
            & 4701.44 &  10  &   8.39 &  0.14 &  0.04 &   0.02 &  0.09 &  0.17 \\
            & 4705.35 &  40  &   8.38 &  0.16 &  0.05 &   0.07 &  0.11 &  0.21 \\
            & 4890.85 &  12  &   8.44 &  0.14 &  0.06 &   0.02 &  0.08 &  0.17 \\
            & 4906.83 &  19  &   8.42 &  0.16 &  0.06 &   0.04 &  0.09 &  0.20 \\
            & 4941.07 &  19  &   8.43 &  0.16 &  0.05 &   0.04 &  0.10 &  0.20 \\
            & 4943.00 &  27  &   8.41 &  0.17 &  0.05 &   0.06 &  0.11 &  0.22 \\
            & 4955.71 &   8  &   8.42 &  0.13 &  0.05 &   0.02 &  0.08 &  0.16 \\
            & 5190.49 &   6  &   8.49 &  0.14 &  0.05 &   0.01 &  0.08 &  0.17 \\
            & 5206.64 &  16  &   8.37 &  0.15 &  0.05 &   0.03 &  0.10 &  0.19 \\
            &&&&&&&&\\

{\bf Ne I}  &  5852.49  &  12 &  8.27 &  0.06 &  0.00  &  0.01 &  0.05 &  0.08  \\ 
            &  6143.06  &  35 &  8.19 &  0.05 &  0.01  &  0.01 &  0.04 &  0.07  \\
            &  6163.59  &  10 &  8.29 &  0.04 &  0.00  &  0.00 &  0.04 &  0.06  \\
            &  6217.28  &   7 &  8.29 &  0.05 &  0.00  &  0.01 &  0.05 &  0.07  \\
            &  6334.43  &  30 &  8.36 &  0.05 &  0.01  &  0.01 &  0.04 &  0.07  \\
            &  6382.99  &  25 &  8.30 &  0.04 &  0.01  &  0.01 &  0.04 &  0.06  \\
            &  6402.25  &  65 &  8.19 &  0.05 &  0.00  &  0.04 &  0.05 &  0.08  \\
            &  6506.53  &  30 &  8.22 &  0.06 &  0.00  &  0.02 &  0.06 &  0.09  \\
            &&&&&&&&\\ 

{\bf Mg II}  & 4481.20 & 160   & 7.60 & 0.06 & 0.01 &  0.16&  0.06 &  0.18 \\ 
             &&&&&&&&\\

{\bf Al III}& 4479.86 &  46   & 6.28 & 0.05 &  0.02 &  0.06 &  0.03 &  0.09 \\ 
            & 4512.54 &  41   & 6.33 & 0.04 &  0.01 &  0.11 &  0.02 &  0.12 \\
            & 4529.04 &  66   & 6.30 & 0.03 &  0.01 &  0.11 &  0.01 &  0.11 \\
            &&&&&&&&\\

{\bf Si II}  &  5056.15  &  33   & 7.04 & 0.12 &  0.03 &  0.04 &  0.09 &  0.16  \\ 
             &  6371.37  &  37   & 7.34 & 0.12 &  0.01 &  0.07 &  0.11 &  0.18  \\
{\bf Si III} &  4567.84  &  97   & 7.26 & 0.11 &  0.04 &  0.20 &  0.07 &  0.24  \\
             &  4574.76  &  65   & 7.28 & 0.12 &  0.04 &  0.14 &  0.07 &  0.20  \\
             &  4813.33  &  22   & 6.94 & 0.12 &  0.04 &  0.04 &  0.08 &  0.15  \\
             &  4819.77  &  32   & 7.06 & 0.13 &  0.04 &  0.04 &  0.08 &  0.16  \\
             &  4829.07  &  33   & 6.89 & 0.12 &  0.03 &  0.04 &  0.08 &  0.15  \\
             &  5739.73  &  69   & 7.32 & 0.14 &  0.04 &  0.13 &  0.08 &  0.21  \\
            &&&&&&&&\\ 
\hline
\end{tabular}
\end{center}
\end{table*}

\newpage

\begin{table*}[h]
\begin{center}
\begin{tabular}{lccc|ccccc} 
\multicolumn {8}{l} {continued...}\\
\\
{\bf Fe III}  & 4310.35  &  13  &  7.18 & 0.04 & 0.02 & 0.06 & 0.00 & 0.07  \\ 
              & 4419.60  &  34  &  7.25 & 0.03 & 0.05 & 0.14 & 0.01 & 0.15  \\
              & 5063.42  &   9  &  7.26 & 0.03 & 0.04 & 0.02 & 0.01 & 0.05  \\
              & 5086.70  &  18  &  7.34 & 0.03 & 0.04 & 0.07 & 0.01 & 0.09  \\
              & 5127.51  &  50  &  7.31 & 0.04 & 0.05 & 0.10 & 0.00 & 0.12  \\
              & 5156.11  &  40  &  7.33 & 0.06 & 0.04 & 0.17 & 0.01 & 0.18  \\
              & 5193.91  &  12  &  7.33 & 0.03 & 0.04 & 0.04 & 0.01 & 0.06  \\
              & 5218.10  &   5  &  7.39 & 0.11 & 0.04 & 0.01 & 0.07 & 0.14  \\
              & 5235.66  &  14  &  7.25 & 0.11 & 0.06 & 0.06 & 0.05 & 0.15  \\
              & 5243.31  &  30  &  7.26 & 0.13 & 0.06 & 0.13 & 0.07 & 0.21  \\
              & 5276.48  &  16  &  7.22 & 0.11 & 0.06 & 0.07 & 0.06 & 0.16  \\
              & 5282.30  &  20  &  7.26 & 0.12 & 0.06 & 0.09 & 0.06 & 0.17  \\
              & 5284.83  &   5  &  7.29 & 0.07 & 0.03 & 0.02 & 0.03 & 0.08  \\
              & 5298.11  &   4  &  7.29 & 0.08 & 0.02 & 0.01 & 0.04 & 0.09  \\
              & 5299.93  &  14  &  7.32 & 0.11 & 0.05 & 0.06 & 0.06 & 0.15  \\
              & 5302.60  &  14  &  7.23 & 0.12 & 0.05 & 0.04 & 0.06 & 0.15  \\
              & 5306.76  &   9  &  7.23 & 0.10 & 0.05 & 0.03 & 0.05 & 0.13  \\
              & 5460.80  &   9  &  7.23 & 0.09 & 0.05 & 0.02 & 0.04 & 0.11  \\
              & 5485.52  &  10  &  7.23 & 0.09 & 0.05 & 0.02 & 0.04 & 0.11  \\
              & 5573.42  &  11  &  7.20 & 0.09 & 0.06 & 0.04 & 0.04 & 0.12  \\
              & 5833.94  &  26  &  7.55 & 0.10 & 0.04 & 0.17 & 0.06 & 0.21  \\
              & 5854.62  &   6  &  7.34 & 0.11 & 0.04 & 0.04 & 0.06 & 0.14  \\
              & 6032.64  &  14  &  7.46 & 0.02 & 0.02 & 0.05 & 0.01 & 0.06  \\
              & 6036.55  &   6  &  7.37 & 0.10 & 0.05 & 0.04 & 0.05 & 0.13  \\
    
\hline
\end{tabular}
\end{center}
\end{table*}

\end{document}